\documentclass[a4paper,fleqn,usenatbib]{mnras}

\usepackage{txfonts}

\usepackage[T1]{fontenc}
\usepackage{ae,aecompl}
\usepackage{graphics}

\usepackage{psfig}   
\usepackage{graphicx}
\usepackage{amssymb}
\usepackage{subfigure}

\title[White Dwarf planets in clusters]{White dwarf planets in star clusters: gravitational scattering versus mass-loss effects}

\author[R.~J.~Parker \& D.~Veras]{
  Richard J.~Parker\thanks{E-mail: R.Parker@sheffield.ac.uk}\thanks{Royal Society Dorothy Hodgkin Fellow}$^1$ and Dimitri Veras$^{2,3,4}$ \vspace*{0.1cm}\\
   $^1$Astrophysics Research Cluster, School of Mathematical and Physical Sciences, The University of Sheffield, Sheffield, S3 7RH, UK \\
  $^2$Department of Physics, University of Warwick, Coventry, CV4 7AL, UK \\
  $^3$Centre for Exoplanets and Habitability, University of Warwick, Coventry, CV4 7AL, UK \\
  $^4$Centre for Space Domain Awareness, University of Warwick, Coventry, CV4 7AL, UK}

\begin{document}

                             
\pagerange{\pageref{firstpage}--\pageref{lastpage}} \pubyear{2023}

\maketitle

\label{firstpage}

\begin{abstract}
White dwarfs are unique laboratories for understanding the formation, evolution and survivability of planetary systems. Post-main sequence mass-loss will change planetary orbital properties and stir up debris discs, leading to the observed pollution of white dwarf atmospheres. However, to date, very few studies have investigated the impact of the stellar birth environment on white dwarf planetary systems. In this paper we simulate the evolution of giant planets around white dwarf progenitors from their formation in a star-forming region until 1\,Gyr, when the most massive stars ($>$2\,M$_\odot$) have left the main sequence. Our simulations self-consistently model $N$-body interactions between stars and planets {\it while} stars evolve into white dwarfs within the cluster lifetime. We find that although scattering interactions in dense star-forming regions create free-floating planets, and alter the orbital properties of up to 20\,per cent of the surviving planets, the effects of mass-loss from the star dominate the dynamics. This behaviour is independent of the stellar density of the birth star-forming region, and largely independent of the initial planet orbital properties. Our simulations produce both captured planets around white dwarfs (potentially similar to WD~0806-661~b), and triple systems with white dwarfs and planets (potentially similar to PSR~B1620-26~(AB)~b), and our results yield a population synthesis of giant planets from 1 -- 100~au that may be relevant to Roman, Gaia and JWST observations.  
\end{abstract}


\section{Introduction}

The observations of planets around white dwarfs \citep{Thorsett93,Sigurdsson03,Luhman11,Gaensicke19,Vanderburg20,Blackman21,Zhang24} provide important constraints on their survivability and even habitability \citep{Ramirez16,Whyte24,Becker25,Mullens25} after their host stars have stopped burning hydrogen. 
While only a handful of such planets are currently known, a couple of thousand white dwarf planetary systems have been discovered through the presence of heavy elements in the stars' photospheres, indicating accretion from a debris disc \citep[e.g.][]{vanMaanen17,Zuckerman87,Williams24}; see \cite{Farihi16} and \cite{Xu24} for reviews.

Because these debris discs have radii which are $\sim 1$\,R$_{\odot}$, they are thought to have formed from the breakup of minor bodies \citep[e.g.][]{Jura03} whose orbits became highly eccentric through gravitational instability due to the presence of planets. A likely trigger for this instability is
post-main sequence mass-loss from the star \citep[e.g.][]{Debes02}; see \cite{Veras24a} for a review. 
This accreted material can then be used to infer the composition of the debris discs of exoplanetary systems \citep[e.g.][]{Gaensicke12,Harrison18,Hollands21,Klein21,Doyle23,Rogers24}, even placing strong constraints on the formation and geophysical evolution of planetary systems \citep[e.g.][]{Doyle19,Buchan22,Bonsor23,Rogers25,Bonsor26}.  

However, the typical birth environment of planetary systems is unclear because the type of  star-forming region \citep[in terms of mass, stellar density and number of massive stars whose photoionising radiation destroys protoplanetary discs, e.g.][]{Hollenbach94,Storzer99,Haworth18b,Haworth23} where exoplanet host stars form is unconstrained \citep{Adams10,Parker20,Reiter22}. Massive, dense star-forming regions would likely disrupt planetary systems due to the strong FUV radiation fields from massive stars \citep[e.g.][]{Johnstone98,Scally01,Adams04,Fatuzzo08,Haworth18b,Winter18b,ConchaRamirez19,Nicholson19a,Parker21a,Qiao23,Qiao26}, and the numerous encounters that could alter the orbits of the planets \citep[e.g.][]{Bonnell01b,Adams06,Parker12a,DaffernPowell22,Schoettler24}. These star-forming regions are likely to be long-lived, but also eject stars (and their planetary systems) into the Galactic field.

In contrast, low-mass star-forming regions are likely to be lower density \citep{Pfalzner16}, contain fewer massive stars \citep{Elmegreen06,Parker07,Maschberger08,Weidner09} but also disperse faster into the Galactic field.  

The origin of white dwarf planetary systems is even harder to constrain due to the old ($\sim$Gyrs) ages of the systems, and the low probability of finding a white dwarf planetary system in an open cluster (which would at the very least place lower limits on the initial density and stellar content of the star-forming region).

Due to conservation of angular momentum, we expect that mass loss from the star will lead to an increase in the orbital semimajor axis of any planets, and potentially a change to the eccentricity of the planet(s) \citep[see e.g.][]{Veras16a}. If these changes occur when the star is still within a clustered environment, then in principle orbit alteration due to mass loss could occur simultaneously with disruption from dynamical encounters. 

For these reasons, a theoretical/computational study into the effects of the birth star-forming region on planetary systems that form and evolve around white dwarf progenitors is timely, especially in the context of observational developments in both direct observations of planets around white dwarfs, and inferred planetary systems through observations of pollution.

Previous research has modelled the long-term effects of dynamical interactions on planetary systems in open clusters \cite[e.g.][]{Laughlin98,Malmberg07b,Hao13,Cai17a,Cai19,FlamminiDotti19}, but has not followed the stellar evolution of the stars  self-consistently within the same $N$-body simulation as they leave the main sequence, lose mass, and ultimately become white dwarfs. Previous research on white dwarf planetary systems in stellar clusters focused on the effects on exo-Kuiper Belts \citep{Veras20a} and super-Earths \citep{Stock22}, but modelled the stellar evolution in a post-processing analysis rather than within the $N$-body simulation (albeit in more detail than in this paper).

In this paper, we follow the dynamical evolution of star-forming regions with $N$-body simulations where every star $M_\star < 2.5$\,M$_\odot$ has one orbiting Jupiter-mass planet at the start of the simulation. We track the encounters that occur in the cluster that occur from the star-forming region, and then implement stellar evolution via look-up tables that synchronise with the timestep in the $N$-body integrator.

The paper is organised as follows. In Section ~\ref{methods} we describe the set-up on the $N$-body simulations and the implementation of stellar evolution. We present our results in Section~\ref{results}, we provide a discussion in Section~\ref{discussion} and we conclude in Section~\ref{conclusion}.

\section{Methods}
\label{methods}

\subsection{Stellar and planetary systems}

We select $N_\star = 1000$ stars from a \citet{Maschberger13} stellar initial mass function (IMF) with a probability density function of the form
\begin{equation}
p(M_{\star}) \propto \left(\frac{M_{\star}}{\mu}\right)^{-\alpha}\left(1 + \left(\frac{M_{\star}}{\mu}\right)^{1 - \alpha}\right)^{-\beta}.
\label{maschberger_imf}
\end{equation}
In Eqn.~\ref{maschberger_imf}  $\mu = 0.2$\,M$_\odot$ is the scale parameter, or `peak' of the IMF \citep{Bastian10,Maschberger13}, $\alpha = 2.3$ is the \citet{Salpeter55} power-law exponent for higher mass stars, and $\beta = 1.4$ describes the slope of the IMF for low-mass objects. The resultant IMF has a similar form to the \citet{Kroupa02} IMF, but with a smoother transition betwen the low-mass and high-mass regimes. The open histogram in Fig.~\ref{JC_MFs} shows the inital stellar mass distribution in our simulations. We randomly sample this IMF between 0.08 -- 50\,M$_\odot$.

For stars in the mass range 0.08 -- 2.5\,M$_\odot$ we place a single Jupiter-mass planet in orbit around the star. Whilst only stars more massive than 2\,M$_\odot$ will evolve into white dwarfs during the 1\,Gyr timescale of our simulations, we include planets around lower-mass stars so that we can contrast the effects of dynamical encounters on planets around white dwarf progenitors with those whose host stars remain on the main sequence for the length of the simulation.

In two sets of simulations we randomly select the semimajor axes of the planets from a flat distribution (i.e.\,\,all values are equally likely) between 0.1 and 50\,au. In three further simulation sets we select the semimajor axes from a delta function at 0.1, 5, and 30\,au. In all simulations the planets' eccentricities and inclinations are initially set to zero, which we expect from planet formation in a circumstellar disc.

For simplicity, we do not include stellar binary systems in our simulations. Our planetary systems are binaries, in the sense that every star with a mass $\leq 2.5$\,M$_\odot$ is orbited by a single Jupiter-mass planet. Stars that do not host a planet (i.e.\,\,$M_{\star} > 2.5$\,M$_\odot$) are all single and do not have a stellar (or substellar) companion.

However, we note that in reality the observed multiplicity fraction of young stars is high \citep{Raghavan10,Sana13,DeRosa14,Ward-Duong15}. Planets have been observed on both circumprimary (`S-type') \citep{Bonavita07,Thebault25} and circumbinary (`P-type') orbits \citep{Orosz12a,Orosz12b}, and in principle a planet could orbit a WD-WD binary system, or orbit one of the component WDs in a binary system \citep{Danielski19,Tamanini19,Columba23,Ledda23,Nigioni26}. Stellar binarity adds a further set of dynamical processes in addition to stellar fly-bys in a star-forming region, such as the von~Zeipel-Lidov-Kozai mechanism \citep{vonZeipel10,Lidov62,Kozai62}. Many authors \citep[e.g.][]{Holman99,Malmberg07a,Parker09c,Ballantyne21,Coleman24} have described these secondary dynamical effects in star-forming regions, but as we do not include primordial stellar binaries in our simulations, we do not explore them in this paper.  

\subsection{Cluster initial conditions}

Observations \citep[e.g.][]{Larson81,Gomez93,Larson95,Cartwright04,Sanchez09,Kuhn14,Henshaw16b,Hacar18,Buckner19,Kuhn19} and simulations \citep[e.g.][]{Schmeja06,Girichidis11} show that star-forming regions initially form with a high degree of spatial and kinematic substructure, which is then erased due to dynamical encounters and relaxation \citep[e.g.][]{Goodwin04a,Parker14b}.

An initially unbound star-forming region is likely to first form an association-like complex \citep{Wright14,Wright20} before dissipating into the Galactic field. However, if the star-forming region is bound, the spatial and kinematic substructure is erased \citep{Parker12d,Parker14b} and the region forms a smooth, centrally concentrated cluster, similar to observed open clusters \citep{Hunt24}.

In this work we set up our star-forming regions as bound, primarily to maximise the chances of dynamical encounters occurring to the planetary systems. Dynamical encounters that would affect planetary orbits are less likely if the regions are unbound, although the most of the important dynamical encounters occur within the first few Myr \citep{Parker12a,DaffernPowell22} when the density is relatively high in both bound and unbound star-forming regions. 

We use the box-fractal method \citep{Goodwin04a,Lomax18,DaffernPowell20} to set up substructured star-forming regions. For a full description of the method we refer the reader to \citet{DaffernPowell20} but we briefly restate it here.

We start by defining a cube whose sides have a length $N_{\rm div} = 2$. We place a `root' particle at the centre of the cube, and the cube is then divided into  $N^3_{\rm div}$ sub-cubes, and a `leaf' particle is placed at the centre of each of these sub-cubes.

The probability that each leaf particle will become a root particle is given by $N^{(D - 3)}_{\rm div}$, where $D$ is the fractal dimension. In three dimensions, when $D = 1.6$ few leaf particles become root particles, and the distribution has a highly substructured appearance. When $D = 3.0$ all of the leaf particles themselves become root particles and so the distribution has a smooth appearance.

The final leaf particles have a small amount of noise added to their positions, to prevent the distribution having a gridded appearance. The procedure sometimes produces more particles than are required, and the excess particles are randomly pruned whilst preserving the desired fractal dimension $D$. 

The velocity structure is also determined by the root+leaf particle generation procedure above. The first root star has a velocity drawn from a Gaussian with mean of zero, and then every leaf star inherits this velocity plus an additional random component drawn from the same Gaussian but multiplied by $\left(1/N_{\rm div}\right)^g$, where $g$ is the number of the root+leaf generation the particle was produced in.

This way of generating velocities means that physically close particles have similar velocities, but those that are distant from one another may have very different velocities. This is physically motivated by the \citet{Larson81} relations, where nearby cores/protostars have small velocity dispersions, and distant cores/protostars have larger velocity dispersions. 

We then scale the velocities to the virial ratio of the star-forming region, $\alpha_{\rm vir} = T/|\Omega|$, where $T$ and $|\Omega|$ are the total kinetic and total potential energies of all the objects, respectively. Again, to replicate observations of the virial ratio of cores and proto-stars in star-forming regions \citep[e.g.][]{Foster15}, we set the virial ratio to be subvirial (i.e.\,\,the region is bound), with $\alpha_{\rm vir} = 0.3$.

We set the simulations up with the highest degree of spatial and kinematic substructure, namely $D = 1.6$. We then scale the positions of the stars to the overall radius of the star-forming regions, $r_F$. In most sets of simulations we adopt $r_F = 1$\,pc (again, to maximise the number of dynamical interactions that could feasibly occur to planets), but in one set we adopt $r_F = 5$\,pc. This results in median stellar densities of $10^4$\,M$_\odot$\,pc$^{-3}$ and $10^2$\,M$_\odot$\,pc$^{-3}$, respectively. The highest density we adopt is likely to be the upper limit for stellar densities in star-forming regions that disperse into the Galactic field \citep{Parker14e,Schoettler22}.

For completeness, we also run a set of simulations in which the positions and velocities of the stars are distributed in a \citet{Plummer11} potential, using the prescription in \citet{Aarseth74}. These simulations have a half-mass radius of 0.1\,pc, which results in a similar median local stellar density to the $r_F = 1$\,pc fractals ($10^4$\,M$_\odot$\,pc$^{-3}$). We scale the velocities such that these simulations are in virial equilibrium ($\alpha_{\rm vir} = 0.5$). Plummer spheres are commonly used to model the evolution of star clusters in $N$-body simulations, though their spatial and velocity distributions bear little resemblence to the early stages of star formation. We will describe the differences (and similarities) between the evolution of the Fractal and Plummer simulations in the Appendix. A summary of the simulations is provided in Table~\ref{cluster_sims}.

\begin{table}
\caption[bf]{Summary of simulation set-ups. The columns show the simulation label (with the label extension `-Pl' to denote a Plummer sphere distribution), the initial radius of the star-forming region for the fractals, $r_F$,  or the initial half-mass radius of the Plummer spheres, $r_{1/2}$, the corresponding initial stellar density, $\tilde{\rho}$, and then the semimajor axis distribution of the planetary systems, $f(a_p)$. The first row is the simulation we refer to as the `default' simulation throughout the paper.}
\begin{center}
\begin{tabular}{|c|c|c|c|}
  \hline
Simulation & $r_F$ or $r_{1/2}$ & $\tilde{\rho}$ & $f(a_p)$ \\
  \hline
A &  1\,pc  & $10^4$\,M$_\odot$\,pc$^{-3}$ & Uniform from 0.1 -- 50\,au \\
  \hline
B &  5\,pc  & $10^2$\,M$_\odot$\,pc$^{-3}$ & Uniform from 0.1 -- 50\,au \\
  \hline
C &  1\,pc  & $10^4$\,M$_\odot$\,pc$^{-3}$ & Delta function at 0.1\,au \\
D &  1\,pc  & $10^4$\,M$_\odot$\,pc$^{-3}$ & Delta function at 1\,au \\
E &  1\,pc  & $10^4$\,M$_\odot$\,pc$^{-3}$ & Delta function at 30\,au \\
\hline
A-Pl & 0.1\,pc & $10^4$\,M$_\odot$\,pc$^{-3}$  & Uniform from 0.1 -- 50\,au \\
E-Pl & 0.1\,pc & $10^4$\,M$_\odot$\,pc$^{-3}$  &  Delta function at 30\,au \\
\hline
\end{tabular}
\end{center}
\label{cluster_sims}
\end{table}

\subsection{Dynamical evolution}

We evolve the star-forming regions as $N$-body simulations for 1\,Gyr using the $4^{\rm th}$-order Hermite integrator \texttt{kira} within the \texttt{Starlab} environment \citep{Zwart99,Zwart01}. Snapshot data are output every 10\,Myr, though close encounters ($<1000$\,au) between stars and planets are recorded continuously. Any change in the semimajor axis or eccentricity of the planet due to encounters is recorded. If a planet is ejected from its host star, it remains as a free-floating planet in the cluster, unless it is subsequently captured by a star.

In order to gauge the level of stochasticity in the dynamical evolution of the star-forming regions, we run ten versions of each set of initial conditions, identical apart from the random number seed used to initialise the masses, positions and velocities of the stars and planets.   We do not impose an external Galactic tidal field on the star-forming regions, though these fields may be important for the long-term dynamical evolution of post-main sequence planetary systems \citep{Veras14,Bonsor15,Veras16b,OConnor23a,Pham24}.

In isolation, the gravitational field of a single planetary system changes when the parent star loses mass during the giant branch phases of stellar evolution. If the system consists of just one planet and the star, and if the mass loss is isotropic, then the planet's semimajor axis $a$, eccentricity $e$ and argument of periastron will all change \citep{Omarov62,Hadjidemetriou63}. However, the change in eccentricity will be small unless the mass loss is particularly quick and strong, and/or if the planet is far away from the star \citep{Veras11}.

The limit of small changes in eccentricity due to mass loss is referred to as the `adiabatic' limit. In this limit, equation 17 of \cite{Veras11} and equation 19 of \cite{Veras14} reveal that the maximum change in eccentricity due to isotropic mass loss is

\begin{equation}
{\rm max} \left( \Delta e \right) \approx 0.05 
\left( 
\frac{{\rm max}\left(\dot{M}_{\star} \right)}{10^{-5}{\rm M_{\odot}}~{\rm yr}^{-1}}
\right)
\left(
\frac{a}{10^3~{\rm au}}
\right)^{3/2}
\left(
\frac{M_{\star}}{1~{\rm M_{\odot}}}
\right)^{-3/2}
.
\label{deltae}
\end{equation}

\noindent{}Hence, for planets within about 100~au of their parent stars, this eccentricity change is negligible for most types of stars. However, for $a\gtrsim 1000$~au, the change in eccentricity due to mass loss may be non-negligible, and the mass loss may no longer be adiabatic.

Our numerical implementation detects all changes in eccentricity that are induced by gravitational perturbations, but does not model the small changes in eccentricity due to stellar mass loss\footnote{Implementing a scheme that would is well beyond the scope of this project, and may not be computationally feasible given the necessary sub-step splicing involved \citep{Veras13a,Mustill18}.}. This neglect does not change our overall results in any substantial way, potentially only affecting the very few systems where planets' orbits are expanded by orders of magnitude.

We have verified that our code does correctly increase our planets' semimajor axes in the adiabatic limit. We are able to make that verification because there is an exact, time-independent solution for orbit expansion in this limit:

\begin{equation}
a_{\rm final} =  a_{\rm initial} 
                 \left( 
                 \frac{M_{\star,{\rm initial}}}
                 {M_{\star,{\rm final}}}
                 \right).
\label{eq:sma_evol}
\end{equation}
(In the two-body problem with isotropic mass loss from one of the bodies, the specific angular momentum of the form $\sqrt{GM_{\star}a(1-e^2)}$ is conserved \citep{Veras11}. Hence, in the aforementioned adiabatic limit, the semimajor axis increases when the mass decreases and we obtain Eq.~\ref{eq:sma_evol}.)

Therefore, given that most main-sequence stars with $M_{\star} \gtrsim 1M_{\odot}$ will lose $\approx$~ 50-80\,per cent of their mass, a planet's semimajor axis will nearly always increase by a factor of 2-5 in the adiabatic limit.

\begin{figure*}
  \begin{center}
\setlength{\subfigcapskip}{10pt}
\hspace*{-1.5cm} \subfigure[0\,Myr]{\label{dyn_evo-a}\rotatebox{270}{\includegraphics[scale=0.25]{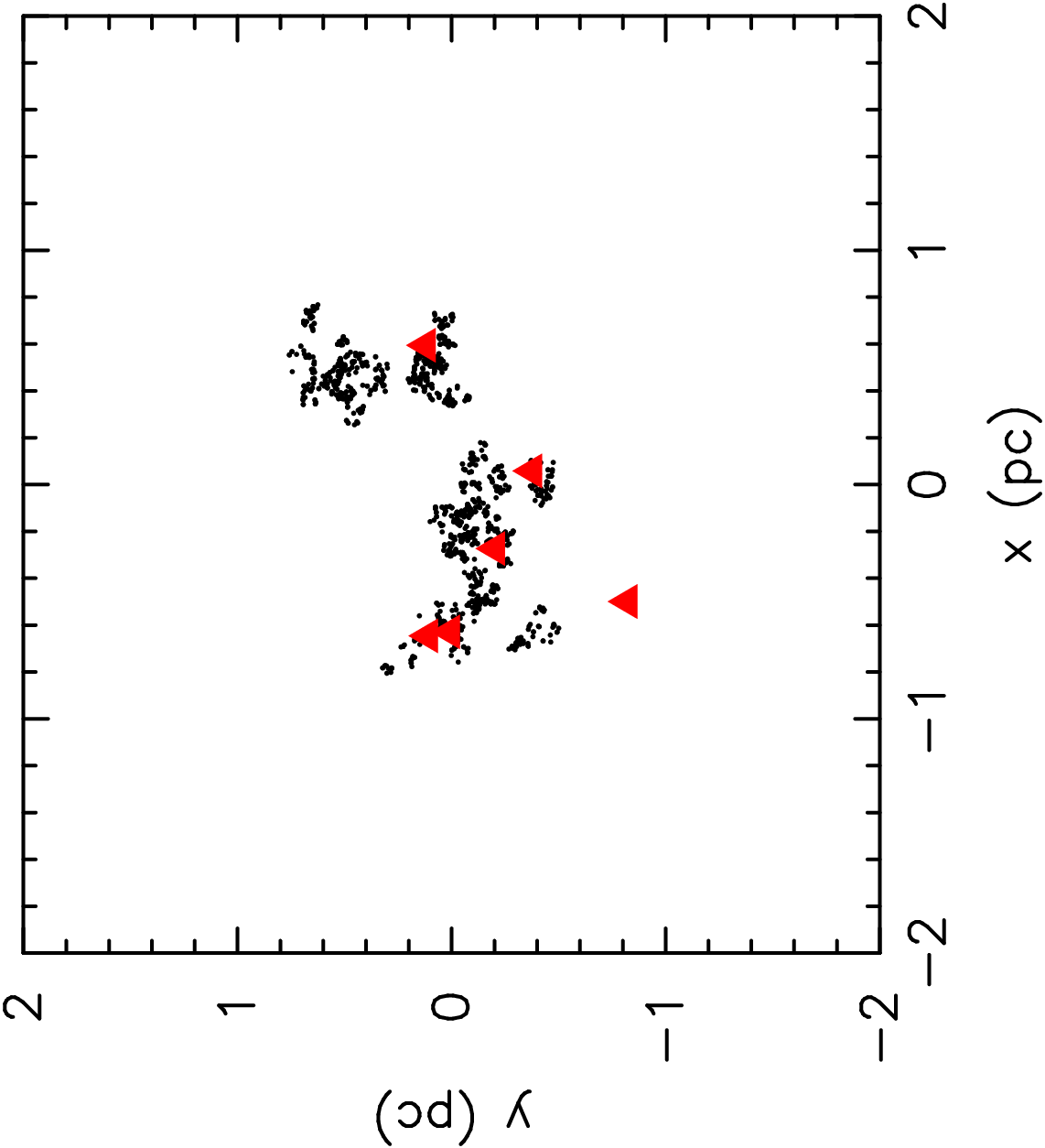}}} 
\hspace*{0.3cm}\subfigure[10\,Myr]{\label{dyn_evo-b}\rotatebox{270}{\includegraphics[scale=0.25]{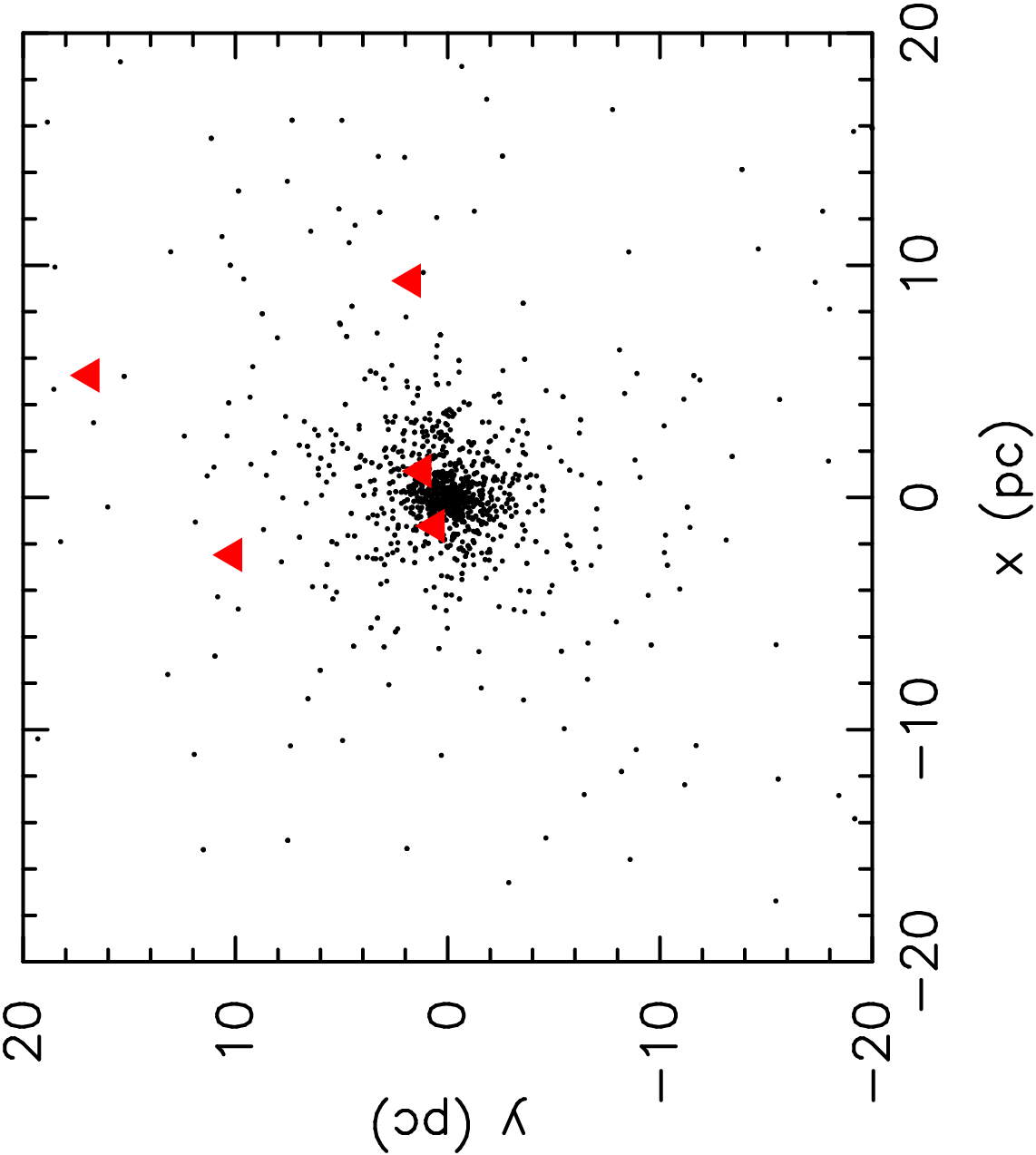}}}
\hspace*{0.3cm}\subfigure[1\,Gyr]{\label{dyn_evo-c}\rotatebox{270}{\includegraphics[scale=0.25]{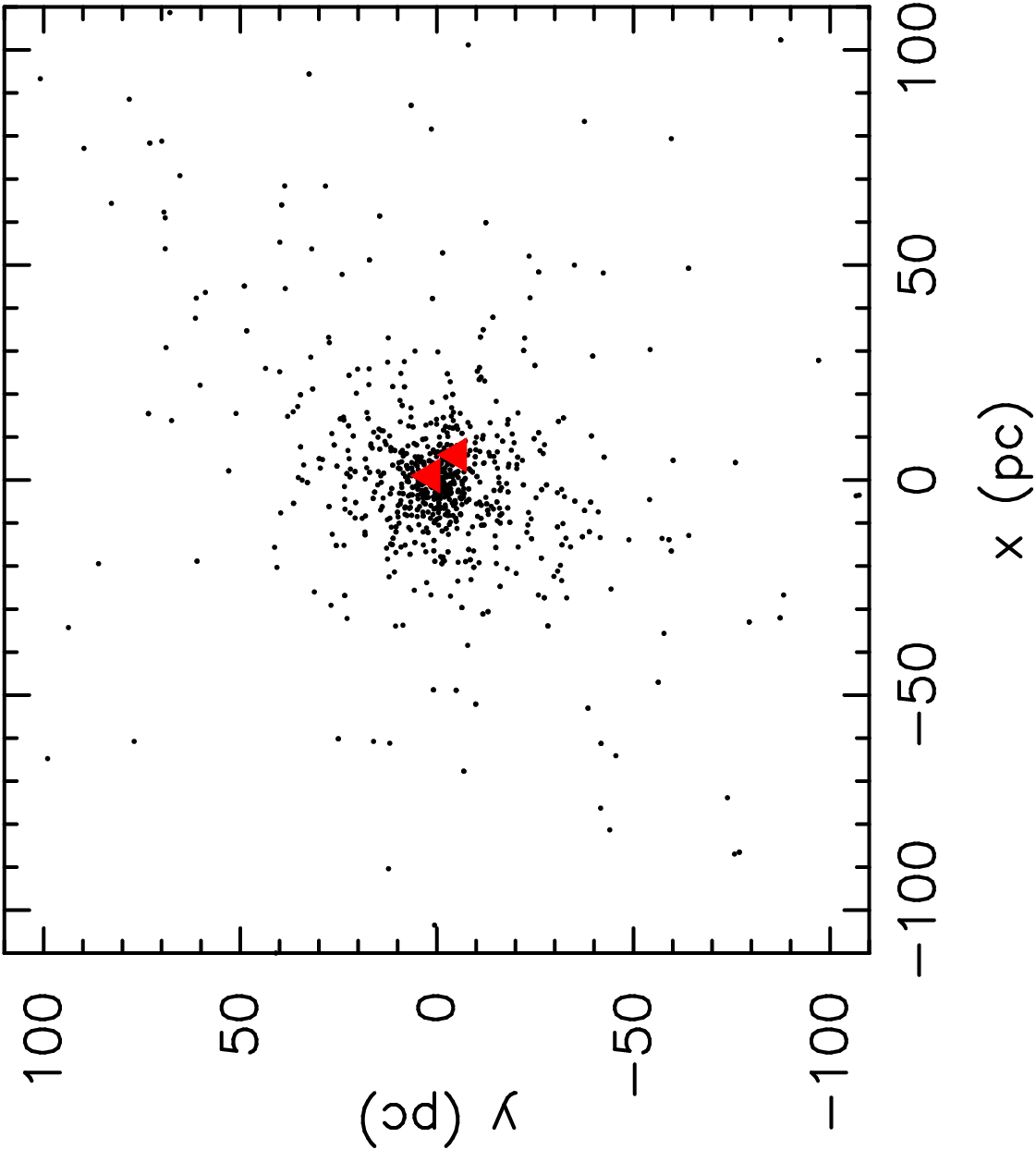}}}
\caption[bf]{Dynamical evolution of one of our default simulations (`A'). We show the positions of stars at 0\,Myr, 10\,Myr and 1\,Gyr. The positions of white dwarf planetary systems where the progenitor has a mass in the range 2 -- 2.5\,M$_\odot$ are shown by the red triangles.}
\label{dyn_evo}
  \end{center}
\end{figure*}

We show three snapshots from a representative simulation from set `A' (our default models) in Fig.~\ref{dyn_evo}. The plots demonstrate the rapid erasure of the initial substructure in the fractal simulations, as they evolve to Plummer-like distributions. We show the positions of white dwarf progenitors with masses between 2 -- 2.5\,M$_\odot$ (we omit other WD progenitors for clarity) that still host their original planets at each time by the red triangles. The number of such systems in this simulation decreases from  6 systems at 0\,Myr to 4 systems after 1\,Gyr, two of which are ejected from the cluster, but the remaining two systems reside in the central regions of the cluster. 

\subsection{Stellar and binary evolution}

We implement stellar and binary evolution using the \texttt{SeBa} look-up tables \citep{Zwart96,Zwart12,Toonen12}, which are incorporated in \texttt{Starlab} and are accessed synchronously with the \texttt{kira} $N$-body integrator  (i.e.\,\,at each timestep).

The \texttt{SeBa} code identifies types of object based on their initial mass (for example, our planetary-mass objects are therefore modelled as giant planets by the code) and is therefore able to follow the evolution of binary systems when one of the component objects (in this case the white dwarf progenitor) begins to lose mass. As there is negligible transfer of mass between the post-main sequence star and the planet, the star and planet merge if the star's radius exceeds the planet's semimajor axis. We consider the planet to be engulfed by the star when such a merger occurs; this process neglects tidal effects, which are likely to be important before engulfment \citep{Mustill12,Madappatt16,Ronco20}.

We set the metallicity of the stars to be Solar; different metallicities can be adopted in  \texttt{SeBa}, similar to other stellar evolution codes \citep[e.g.][]{Tanikawa20,Kamlah22}.

\section{Results}
\label{results}

\subsection{Initial and final mass distributions}

\begin{figure}
\begin{center}
\rotatebox{270}{\includegraphics[scale=0.35]{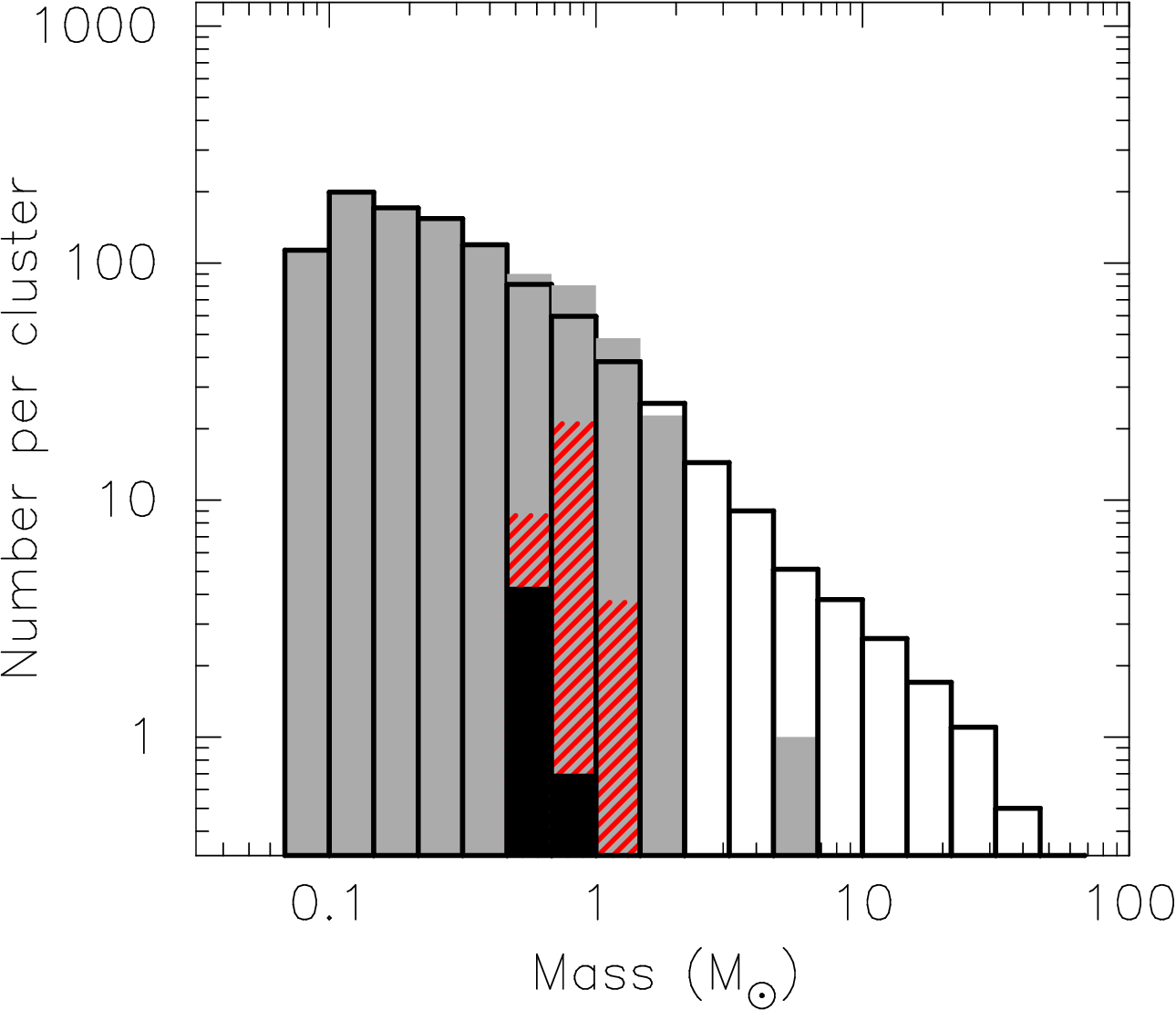}}
\end{center}
\caption[bf]{Mass distributions in our simulations (ten realisations of initial conditions summed together and normalised to numbers per cluster). The open histogram is the initial mass function of the stars. The grey histogram is the final mass function after 1\,Gyr, following stellar evolution. The red hatched histogram is the mass distribution of white dwarfs after 1\,Gyr (this includes stars initially 2.5 -- 8\,M$_\odot$ that did not host planets, as well as stars that have lost their planets). The solid black histogram is the mass distribution of white dwarfs that still host planets after 1\,Gyr.}
\label{JC_MFs}
\end{figure}

We first show the effects of the internal stellar evolution on the mass distribution of stars in the simulations. Fig.~\ref{JC_MFs} shows the average stellar mass distributions in our simulations, where we have summed together ten realisations of the same initial conditions (the simulation set shown are our default set (`A') but the mass functions are the same in all sets).  The initial stellar masses are drawn from a standard \citet{Maschberger13} IMF described in Section~\ref{methods} with the form in Eqn~\ref{maschberger_imf}. This initial mass distribution is shown by the open histogram in Fig.~\ref{JC_MFs}. Following stellar (and binary) evolution, the final mass distribution for all stars/stellar remnants after 1\,Gyr is shown by the grey filled histogram. The red hatched histogram shows the mass distribution of white dwarfs after 1\,Gyr, and the black histogram shows the mass distribution of white dwarfs that host planets after 1\,Gyr. The difference between the mass distribution for all white dwarfs and the mass distribution for the planet hosting white dwarfs is due to the evolution of more massive progenitors ($2.5 - 8$\,M$_\odot$) that did not host planets at the start of the simulations.

\subsection{Encounters}

Due to their kinematic substructure, the initial evolution of the fractal simulations occurs via violent relaxation \citep{LyndenBell67}, which erases the spatial substructure and leads to the formation of a centrally concentrated, smooth stellar cluster (see Fig.~\ref{dyn_evo}). Following this, the cluster evolves and expands due to two-body relaxation.

\begin{figure}
\begin{center}
\rotatebox{270}{\includegraphics[scale=0.35]{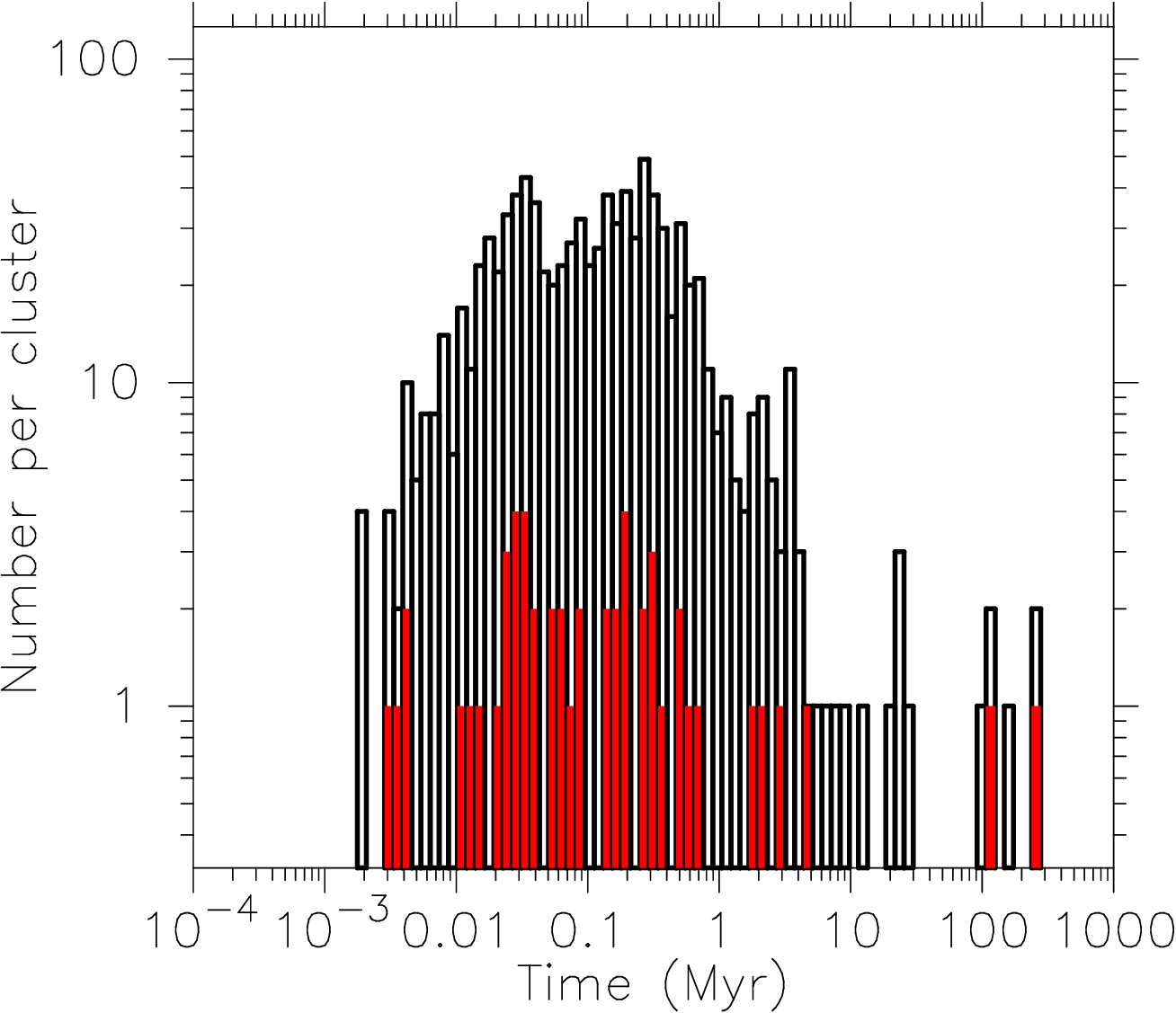}}
\end{center}
\caption[bf]{The number of encounters over time in a representative simulation from our default models (set `A'). The open histogram shows the number of encounters for all stars, and the red histogram shows the number of encounters for stars that will become white dwarfs after leaving the main sequence.  }
\label{fractal_encounters}
\end{figure}

  In Fig.~\ref{fractal_encounters} we show the number of encounters over time as one of our fractal simulations evolves. The open histogram shows the numbers of encounters for all stars, whereas the red filled histogram shows the encounters involving white dwarf progenitors. There is a slight bimodality to the distribution due to an initial spike in encounters as the substructure is erased through violent relaxation, before a second spike caused by the collapse to a centrally concentrated cluster.

This plot also demonstrates that the majority of dynamical encounters occur within the first 10\,Myr of the simulations, well before the first white dwarf progenitor stars leave the main sequence. Despite the very different initial conditions, the similar initial densities mean that the Plummer sphere simulations have a similar encounter rate history (actually slightly fewer encounters per cluster -- see Fig.~\ref{Plummer_encounters} in the Appendix), albeit without the bimodality seen in the fractal simulations, as the Plummer spheres evolve purely due to two-body relaxation. 

\subsection{Destruction of planetary systems before stellar evolution}

The clustered environment is responsible for changing the initial distributions of planetary systems due to (often multiple) encounters with passing stars within the first few Myr. In the dense star-forming regions we simulate, the majority of planetary systems are altered in the first Myr, whereas in the less dense simulation this destruction and alteration of orbits occurs over longer timescales (between 5 -- 10\,Myr).

\begin{figure*}
  \begin{center}
\setlength{\subfigcapskip}{10pt}
\hspace*{-1.5cm} \subfigure[Semimajor axis versus time]{\label{uniform_au_09-a}\rotatebox{270}{\includegraphics[scale=0.35]{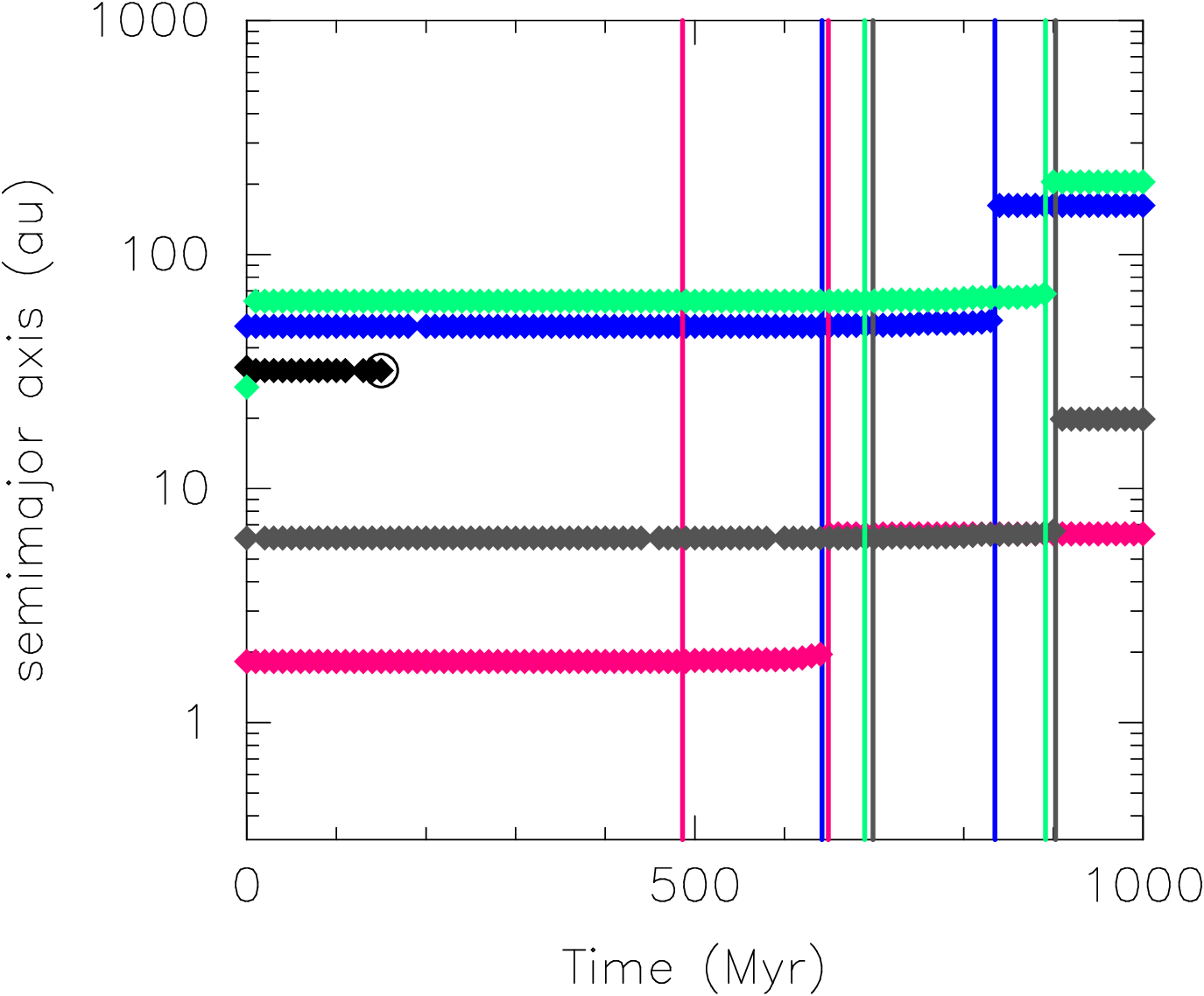}}} 
\hspace*{0.3cm}\subfigure[Eccentricity versus time]{\label{uniform_au_09-b}\rotatebox{270}{\includegraphics[scale=0.35]{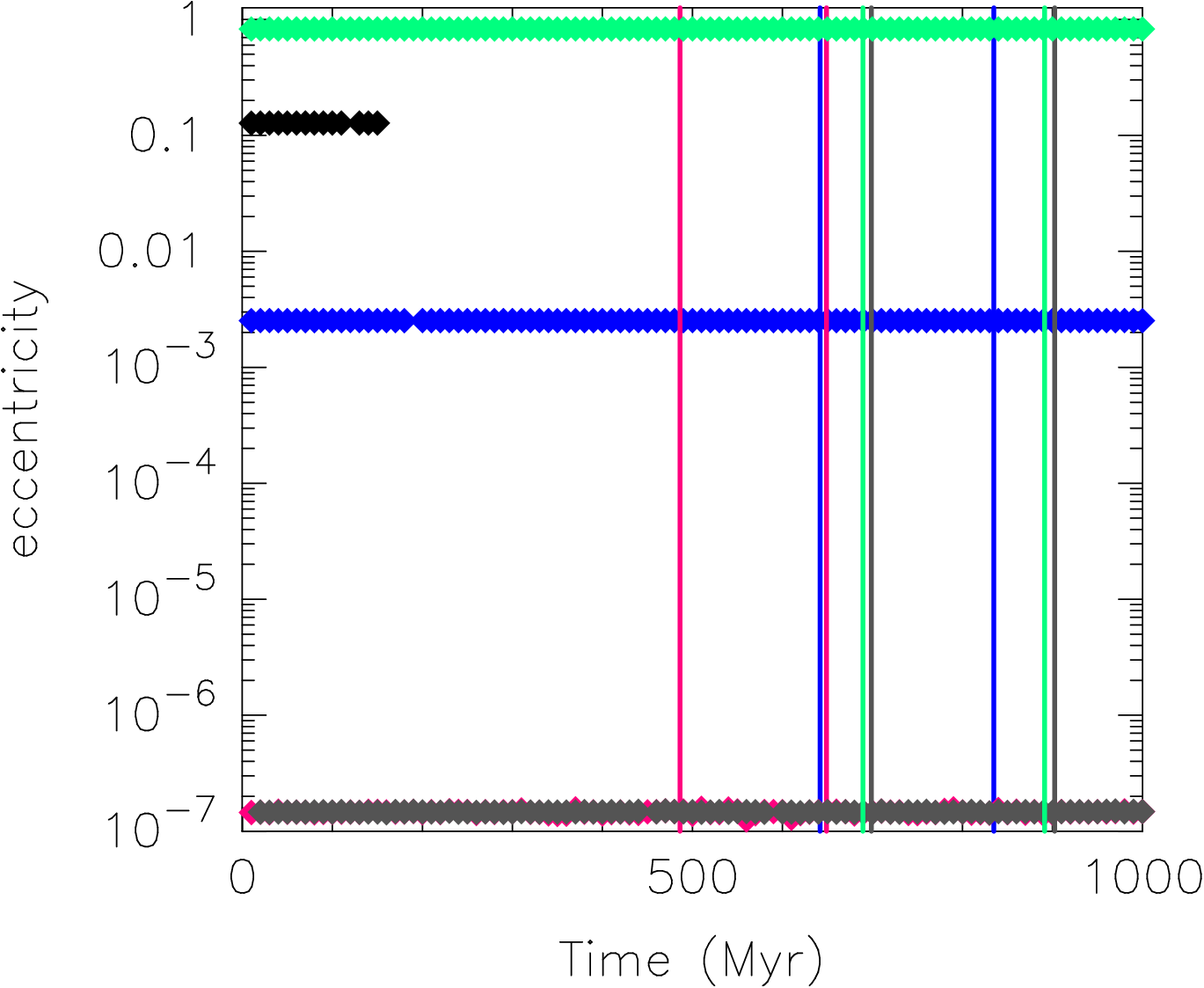}}}

\caption[bf]{The evolution of semimajor axis (panel a) and eccentricity (panel b) with time in one star cluster for planets orbiting stars with initial masses $>$2\,M$_\odot$. The pairs of vertical lines of the same colours as the points indicate the times of the first and last post-main sequence stellar evolution phases. The points represent values taken at snapshots of 10\,Myr intervals during the simulation. The open circle over the black points indicates that the planet has been engulfed by the star. The small gaps that are visible in the sequence of the black, blue and dark grey points are the instances of those planetary systems temporarily entering into a triple system with a third object. }
\label{uniform_au_09}
  \end{center}
\end{figure*}

We use our default simulation (`A' -- a dense star-forming region with planet semimajor axes drawn from a uniform distribution between 0.1 and 50\,au) as an illustrative example. Across the ten realisations of these initial conditions, there are 4576 planets in total, 330 of which (7\,per cent) orbit the progenitor stars of white dwarfs (i.e.\,\,between 1 -- 2.5\,M$_\odot$). After 1\,Gyr, 3237 planets remain on bound orbits around their parent stars (a fraction of 0.71), but only 170 planets remain on bound orbits around white dwarf progenitors (a fraction of 0.52).

This high destruction rate is similar to previous calculations \citep[e.g.][]{DaffernPowell22}, and the higher fraction for the white dwarf progenitors is due to the higher numbers of interactions experienced by more massive stars in a star-forming region \citep[e.g.][]{Allison10}. A small number of planets (15 across all ten realisations of the simulation) are engulfed by the star whilst on the red giant branch phase of its stellar evolution.

\subsection{Evolution of orbits}

Before describing the effects of dynamical encounters and mass loss due to stellar evolution on the semimajor axis and eccentricity distributions, we first illustrate the effects on individual systems within a representative star-forming region. 

In Fig.~\ref{uniform_au_09} we show the semimajor axis (panel a) and eccentricity (panel b) over time for planetary systems where the host star has a mass $>2$\,M$_\odot$, so will evolve to a white dwarf within 1\,Gyr. The pairs of coloured vertical lines show the first and last post-main sequence stellar evolution phases, meaning the first vertical line shows the time the star first leaves the main sequence, and the second vertical line is the time the star becomes a white dwarf.

The coloured points show the corresponding semimajor axes and eccentricities over time of planets that are still bound to their host star after 10\,Myr. Each data point represents an output from the $N$-body simulation, which are produced at 10\,Myr intervals. In these dense simulations, up to 50\,per cent of planetary systems can be destroyed to create free-floating planets \citep{DaffernPowell22}. In this particular simulation, which is dense ($10^4$\,M$_\odot$\,pc$^{-3}$) and the planets have semimajor axes in the range 0.1 -- 50\,au, 29\,per cent of \emph{all} planets become free-floating, but 48\,per cent of the planets around stars that will evolve into white dwarfs become free-floating. This is because stars with a higher mass tend to have more encounters than stars with lower masses in the clusters \citep{Parker23d}.

One system, shown by the green points, has its orbit perturbed by a dynamical encounter after 10\,Myr, which increases its semimajor axis from 27\,au to 63\,au. This interaction also increases the eccentricity from zero to 0.83, though this planet's orbit then remains stable until the star becomes a white dwarf after 891\,Myr. The star has an initial mass of 2.11\,M$_\odot$, and leaves the main sequence after 689\,Myr. After 891\,Myr it becomes a 0.65\,M$_\odot$ white dwarf and the planet's semimajor axis increases from 68\,au to 204\,au. The eccentricity remains at 0.83.

Most planets that remain in orbit around white dwarfs only experience a significant change once the star leaves the main sequence. In Fig.~\ref{uniform_au_09} the system shown by the magenta points is a star with initial mass 2.38\,M$_\odot$ and the planet has an initial semimajor axis of 1.82\,au. The star leaves the main sequence after 486\,Myr, loses mass and becomes a white dwarf after 649\,Myr. The star's mass drops from 2.22\,M$_\odot$ to its final mass of 0.68\,M$_\odot$. The planet's semimajor axis suddenly increases from 1.95\,au to 6.4\,au, although the eccentricity of the planet does not change.

A small number of planets are engulfed by their star as it evolves. In the simulations with planets drawn from a flat distribution between 0.1 -- 50\,au, 4.5\,per cent of the planets orbiting white dwarf progenitors are engulfed by the end of the simulation.  This engulfment mainly happens to planets on very close ($<1$\,au) orbits, but in the simulation we show in Fig.~\ref{uniform_au_09} the system shown by the black points interacts with another star (and actually briefly forms a triple system with that star, seen by the slight gap in the points around 120\,Myr) and the interaction pushes the planet towards  the star and it is engulfed after 150\,Myr, long before the (2.07\,M$_\odot$) star leaves the main sequence.

\begin{figure*}
  \begin{center}
\setlength{\subfigcapskip}{10pt}
\hspace*{-1.5cm} \subfigure[]{\label{sma_uniform-a}\rotatebox{270}{\includegraphics[scale=0.25]{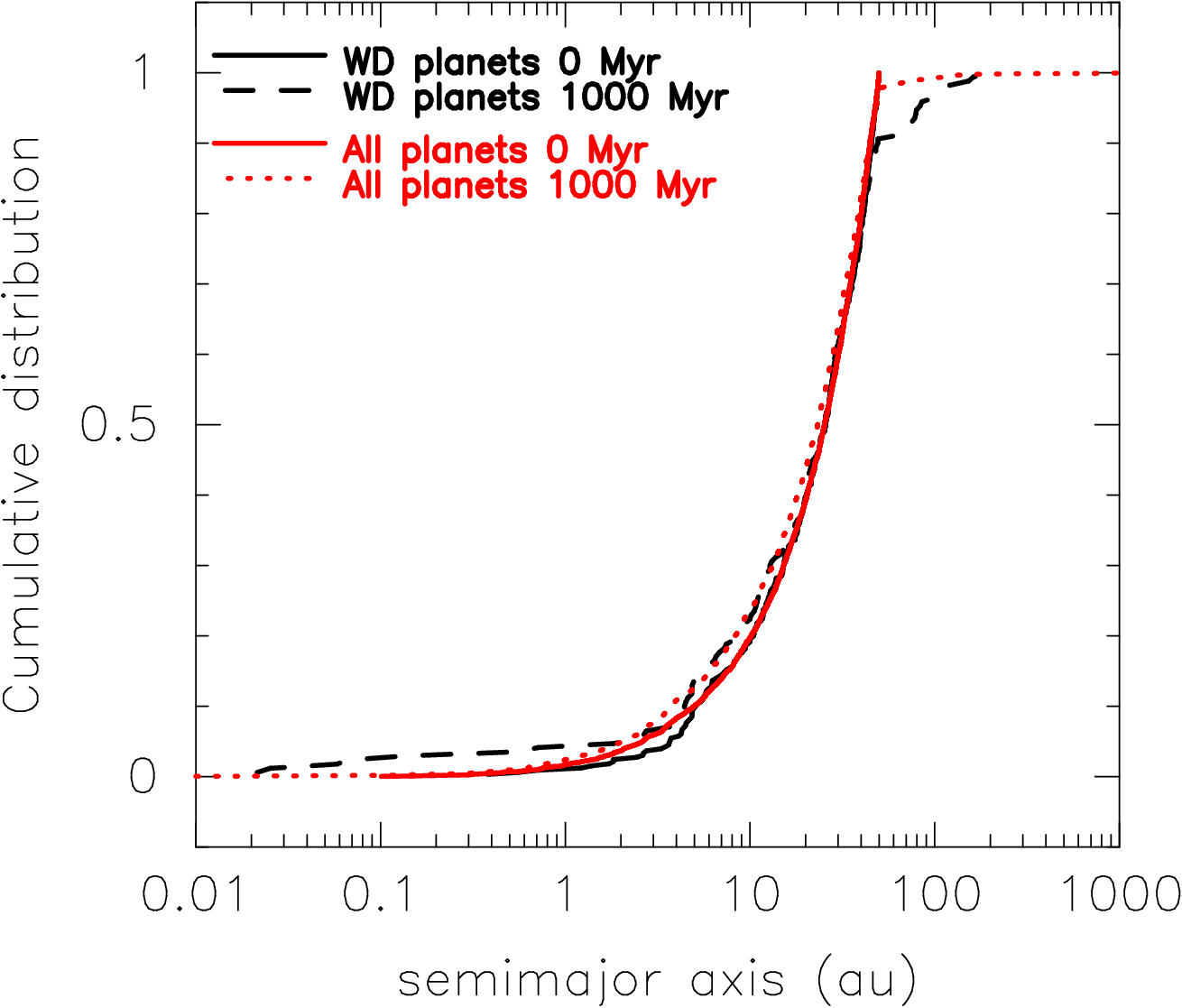}}} 
\hspace*{0.3cm}\subfigure[WD planets]{\label{sma_uniform-b}\rotatebox{270}{\includegraphics[scale=0.25]{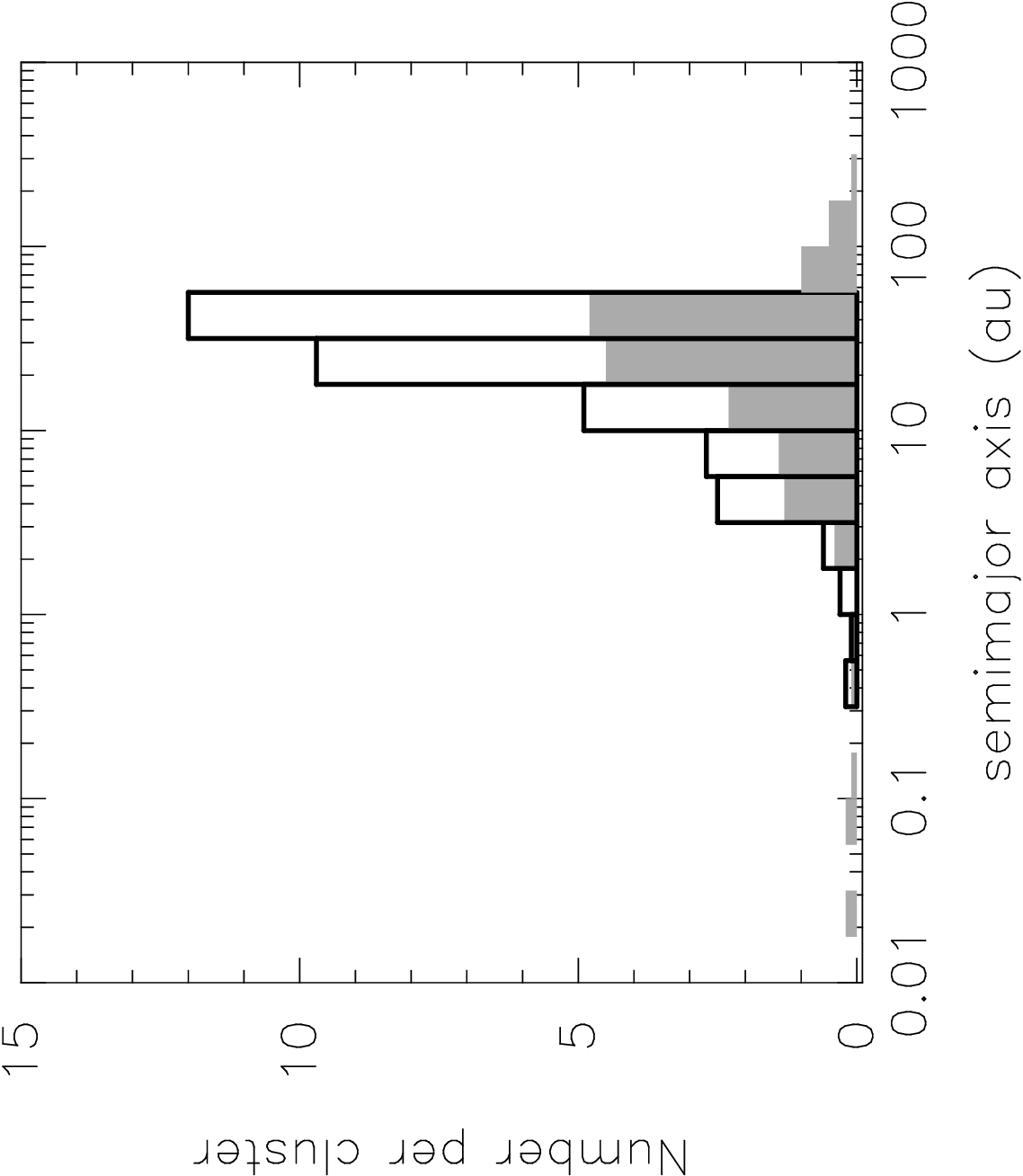}}}
\hspace*{0.3cm}\subfigure[All planets]{\label{sma_uniform-c}\rotatebox{270}{\includegraphics[scale=0.25]{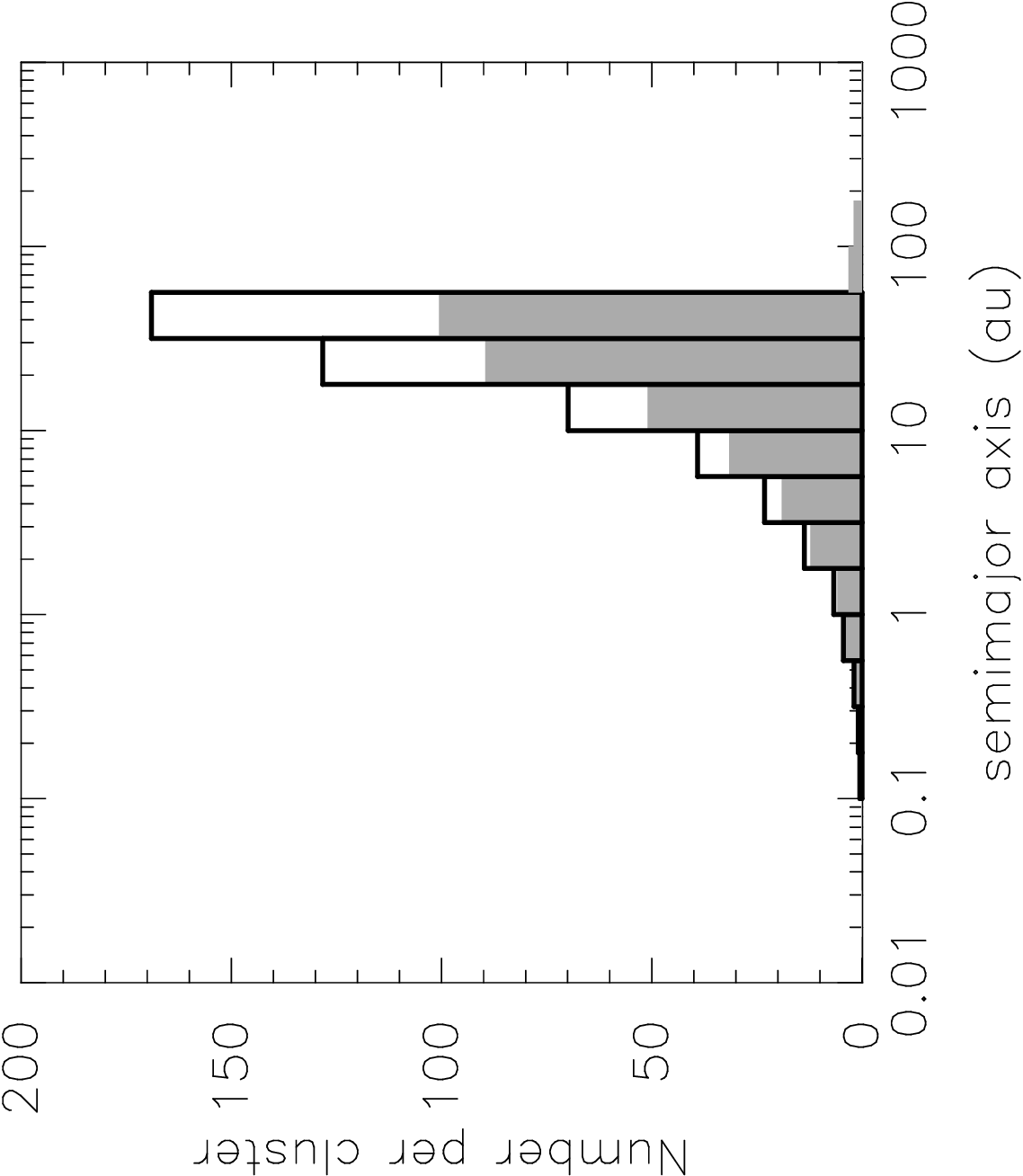}}}

\caption[bf]{Semimajor axes distributions of planets orbiting white dwarf progenitors versus the distributions for all planets in the simulations, for dense star-forming regions ($\tilde{\rho} \sim 10^4$\,M$_\odot$\,pc$^{-3}$) where the initial planet semimajor axes are drawn from a uniform distribution between 0.1 -- 50\,au (simulation set `A'). The final distributions are taken after 1\,Gyr. Panel (a) shows the cumulative distributions of the semimajor axes of all planets (red) and white dwarf planets (black).  The planets shown are only those that orbit their original (parent) star. Panel (b) shows the histogram for the initial (open) and final (grey fill) histograms for the white dwarf planets and panel (c) shows the initial (open) and final (grey fill) histograms for all planets.
}
\label{sma_uniform}
  \end{center}
\end{figure*}

\begin{figure*}
  \begin{center}
\setlength{\subfigcapskip}{10pt}
\hspace*{-1.5cm} \subfigure[]{\label{sma_uniform_100myr-a}\rotatebox{270}{\includegraphics[scale=0.25]{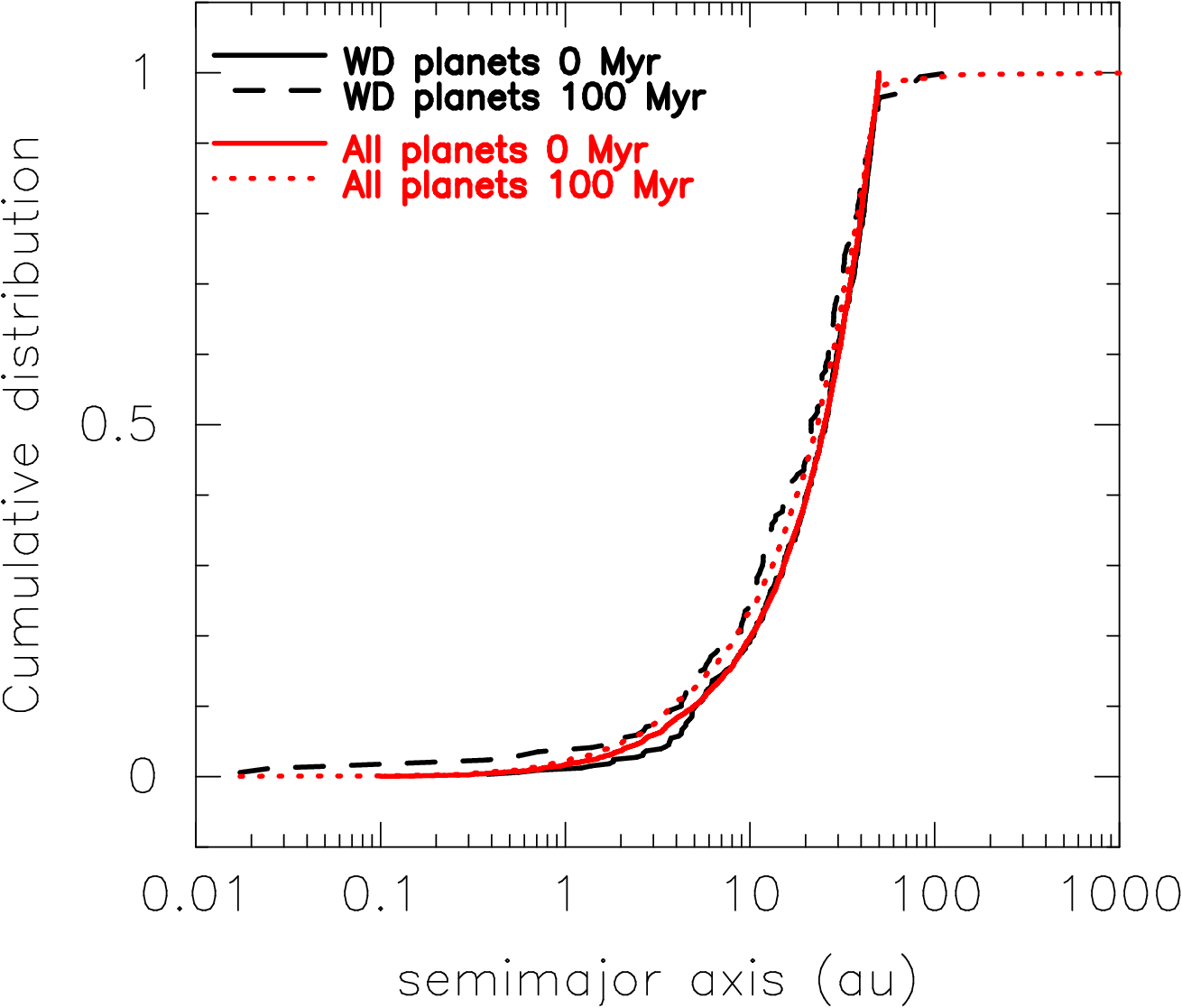}}} 
\hspace*{0.3cm}\subfigure[WD planets]{\label{sma_uniform_100myr-b}\rotatebox{270}{\includegraphics[scale=0.25]{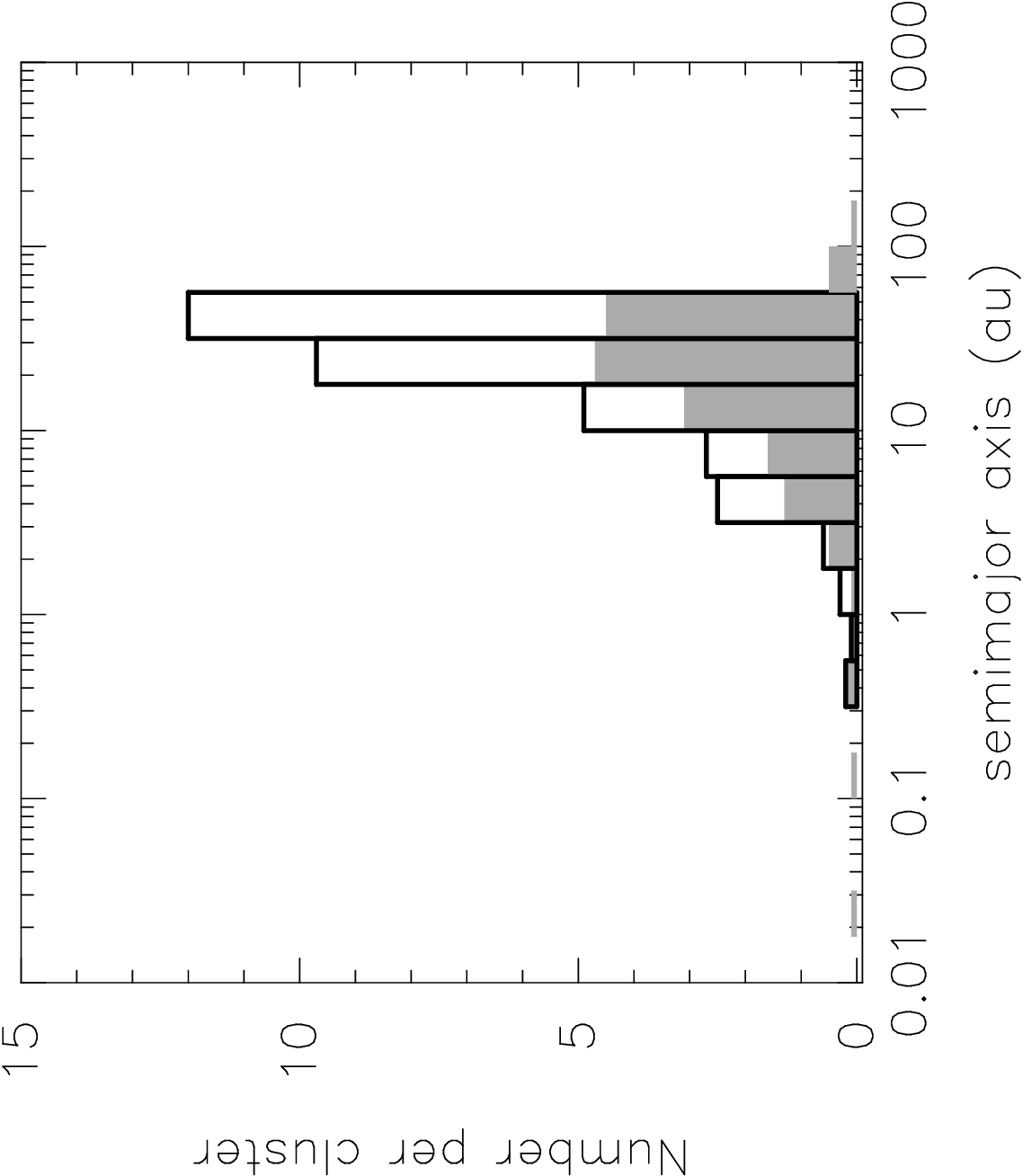}}}
\hspace*{0.3cm}\subfigure[All planets]{\label{sma_uniform_100myr-c}\rotatebox{270}{\includegraphics[scale=0.25]{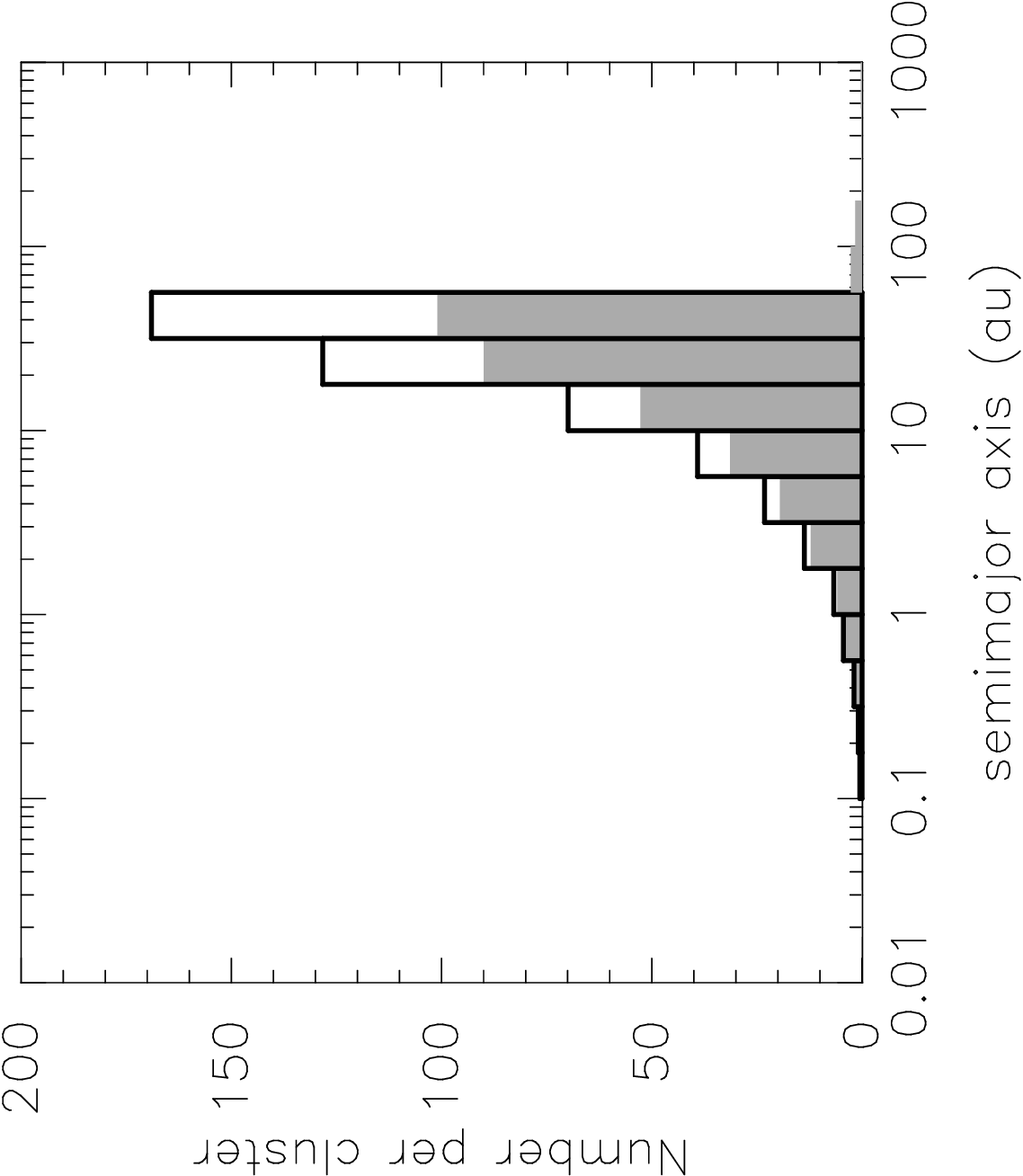}}}

\caption[bf]{As Fig.~\ref{sma_uniform} but where the final semimajor axes distributions are taken after 100\,Myr instead of 1\,Gyr. Semimajor axes distributions of planets orbiting white dwarf progenitors versus the distributions for all planets in the simulations, for dense star-forming regions ($\tilde{\rho} \sim 10^4$\,M$_\odot$\,pc$^{-3}$) where the initial planet semimajor axes are drawn from a uniform distribution between 0.1 -- 50\,au (simulation set `A'). The final distributions are taken after 100\,Myr. Panel (a) shows the cumulative distributions of the semimajor axes of all planets (red) and white dwarf planets (black).  The planets shown are only those that orbit their original (parent) star. Panel (b) shows the histogram for the initial (open) and final (grey fill) histograms for the white dwarf planets and panel (c) shows the initial (open) and final (grey fill) histograms for all planets.
}
\label{sma_uniform_100myr}
  \end{center}
\end{figure*}

\subsubsection{Semimajor axis distributions}

Fig.~\ref{sma_uniform} shows the change in the semimajor axes distributions of the planets in a dense ($10^4$\,M$_\odot$\,pc$^{-3}$) star-forming region, where the initial planetary semimajor axes are drawn from a uniform distribution between 0.1 -- 50\,au (simulation set `A' in Table~\ref{cluster_sims}). In this figure we just show the planets that are still orbiting their original (parent) star.

\begin{figure*}
  \begin{center}
\setlength{\subfigcapskip}{10pt}
\hspace*{-1.5cm} \subfigure[]{\label{sma_uniform_lowrho-a}\rotatebox{270}{\includegraphics[scale=0.25]{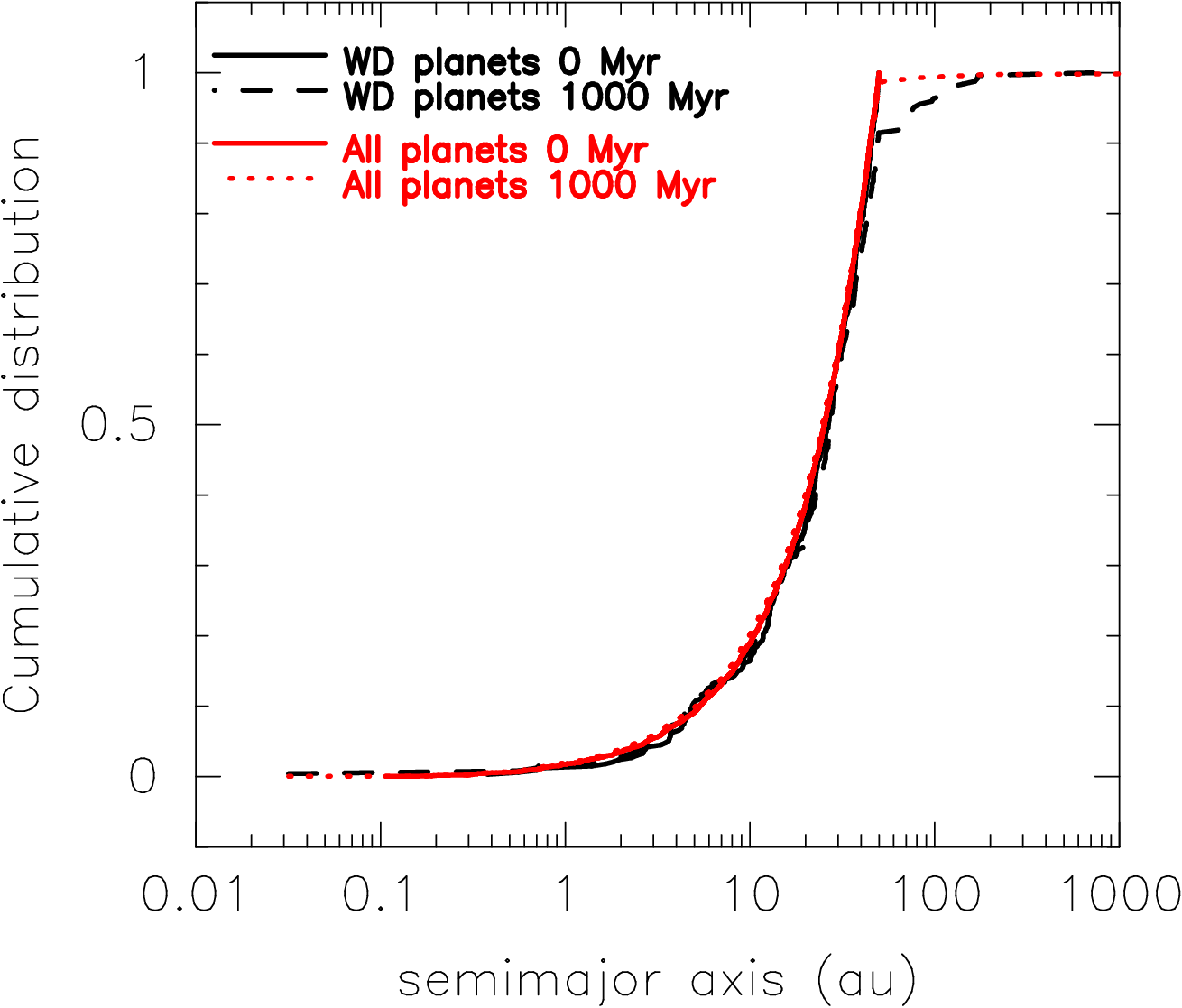}}} 
\hspace*{0.3cm}\subfigure[WD planets]{\label{sma_uniform_lowrho-b}\rotatebox{270}{\includegraphics[scale=0.25]{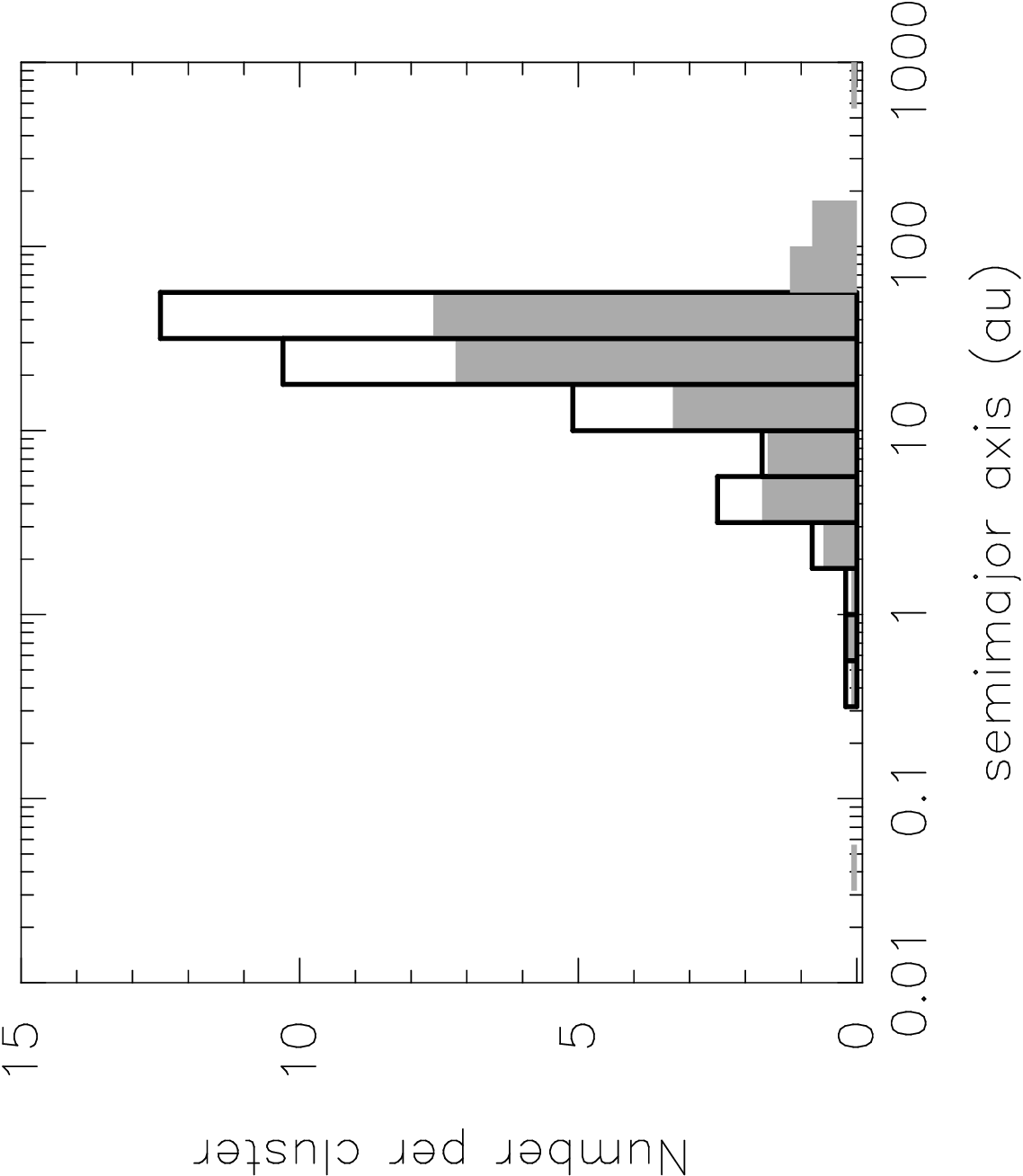}}}
\hspace*{0.3cm}\subfigure[All planets]{\label{sma_uniform_lowrho-c}\rotatebox{270}{\includegraphics[scale=0.25]{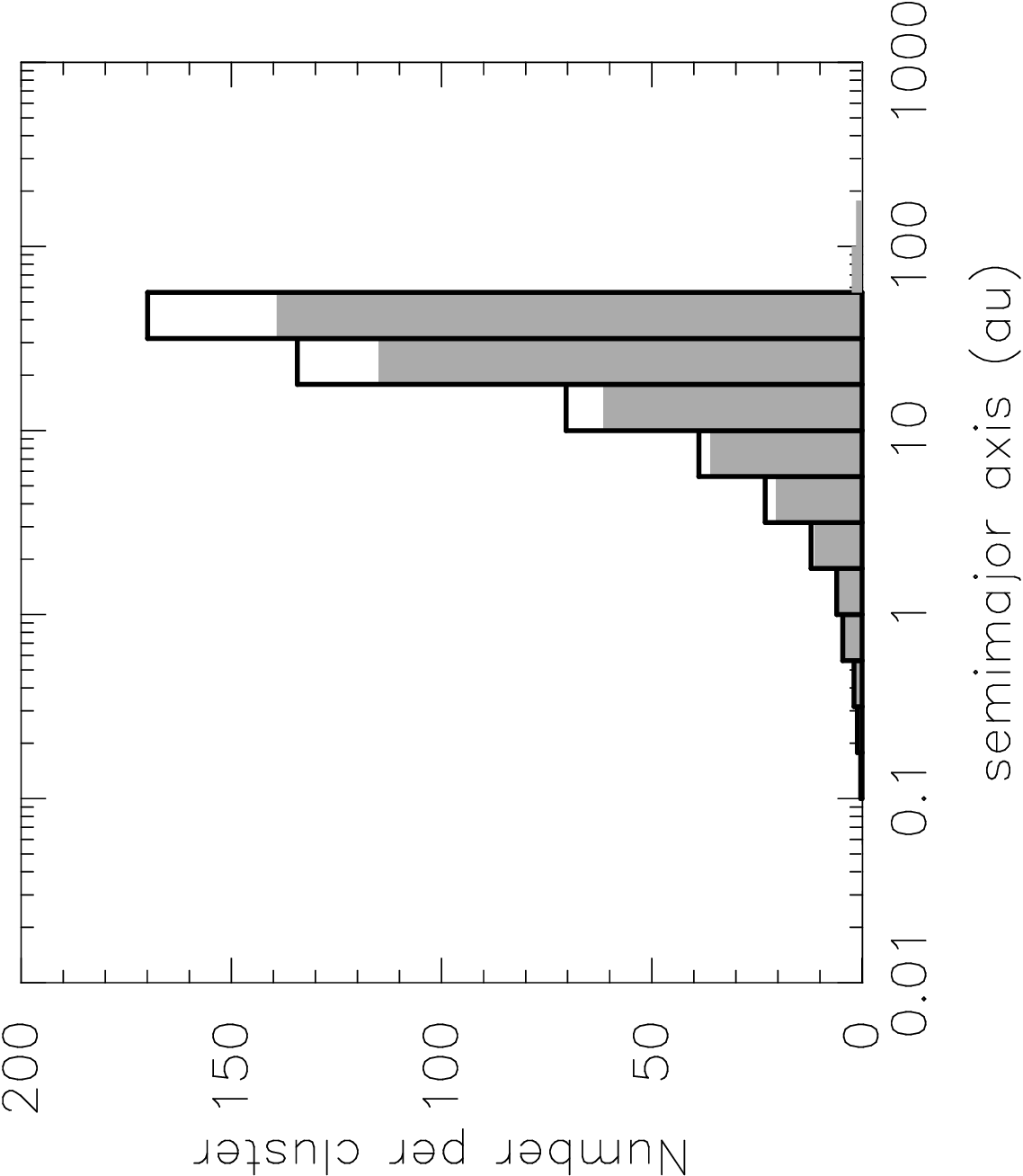}}}

\caption[bf]{Semimajor axes distributions of planets orbiting white dwarf progenitors versus the distributions for all planets in the simulations, for lower-density star-forming regions ($\tilde{\rho} \sim 10^2$\,M$_\odot$\,pc$^{-3}$ -- simulation set `B') where the initial planet semimajor axes are drawn from a uniform distribution between 0.1 -- 50\,au. The final distributions are taken after 1\,Gyr. Panel (a) shows the cumulative distributions of the semimajor axes of all planets (red) and white dwarf planets (black).  The planets shown are only those that orbit their original (parent) star. Panel (b) shows the histogram for the initial (open) and final (grey fill) histograms for the white dwarf planets and panel (c) shows the initial (open) and final (grey fill) histograms for all planets.
}
\label{sma_uniform_lowrho}
  \end{center}
\end{figure*}

\begin{figure*}
  \begin{center}
\setlength{\subfigcapskip}{10pt}
\hspace*{-1.5cm} \subfigure[0.1\,au]{\label{sma_0p1_au-a}\rotatebox{270}{\includegraphics[scale=0.25]{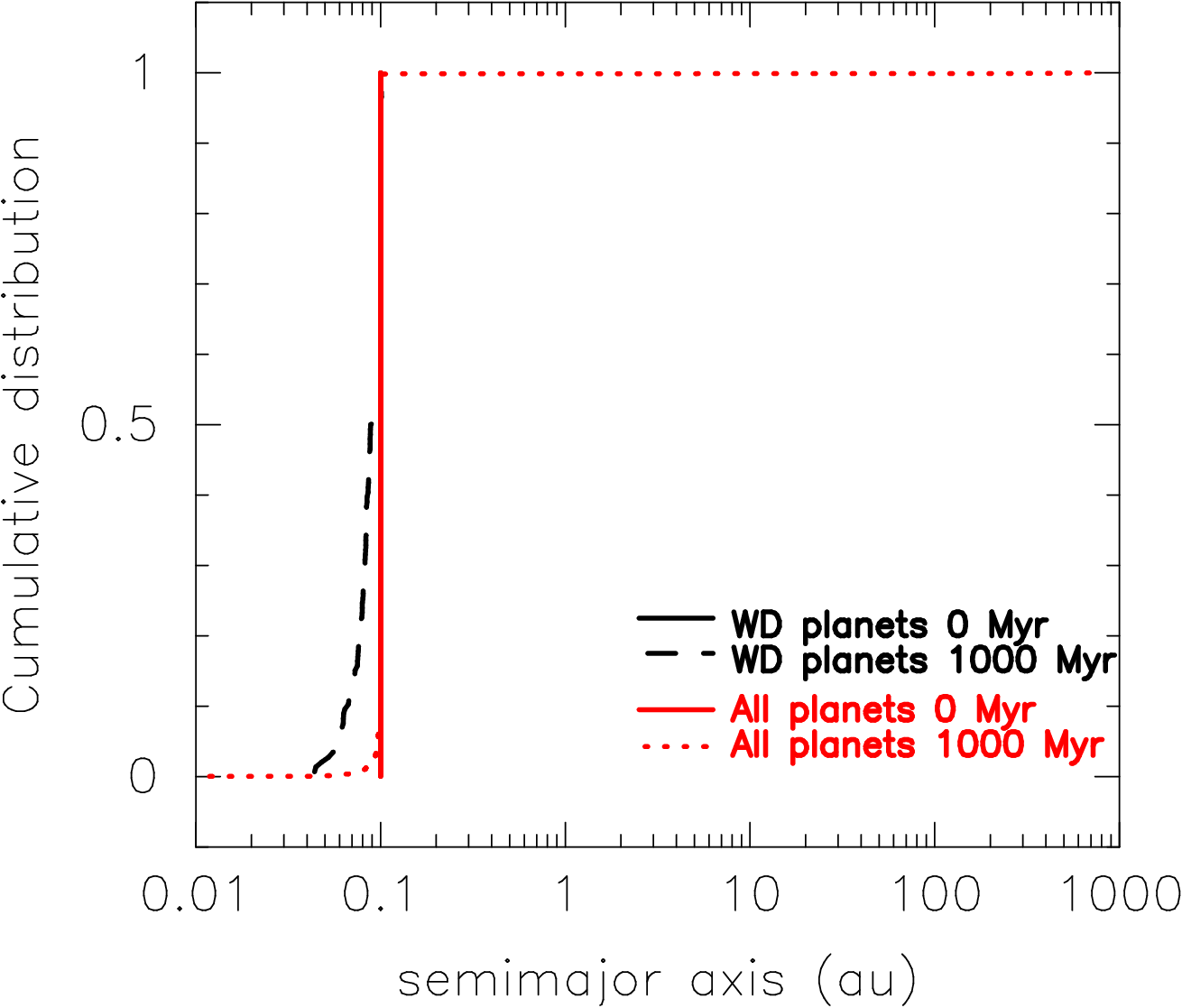}}} 
\hspace*{0.3cm}\subfigure[1\,au]{\label{sma_0p1_au-b}\rotatebox{270}{\includegraphics[scale=0.25]{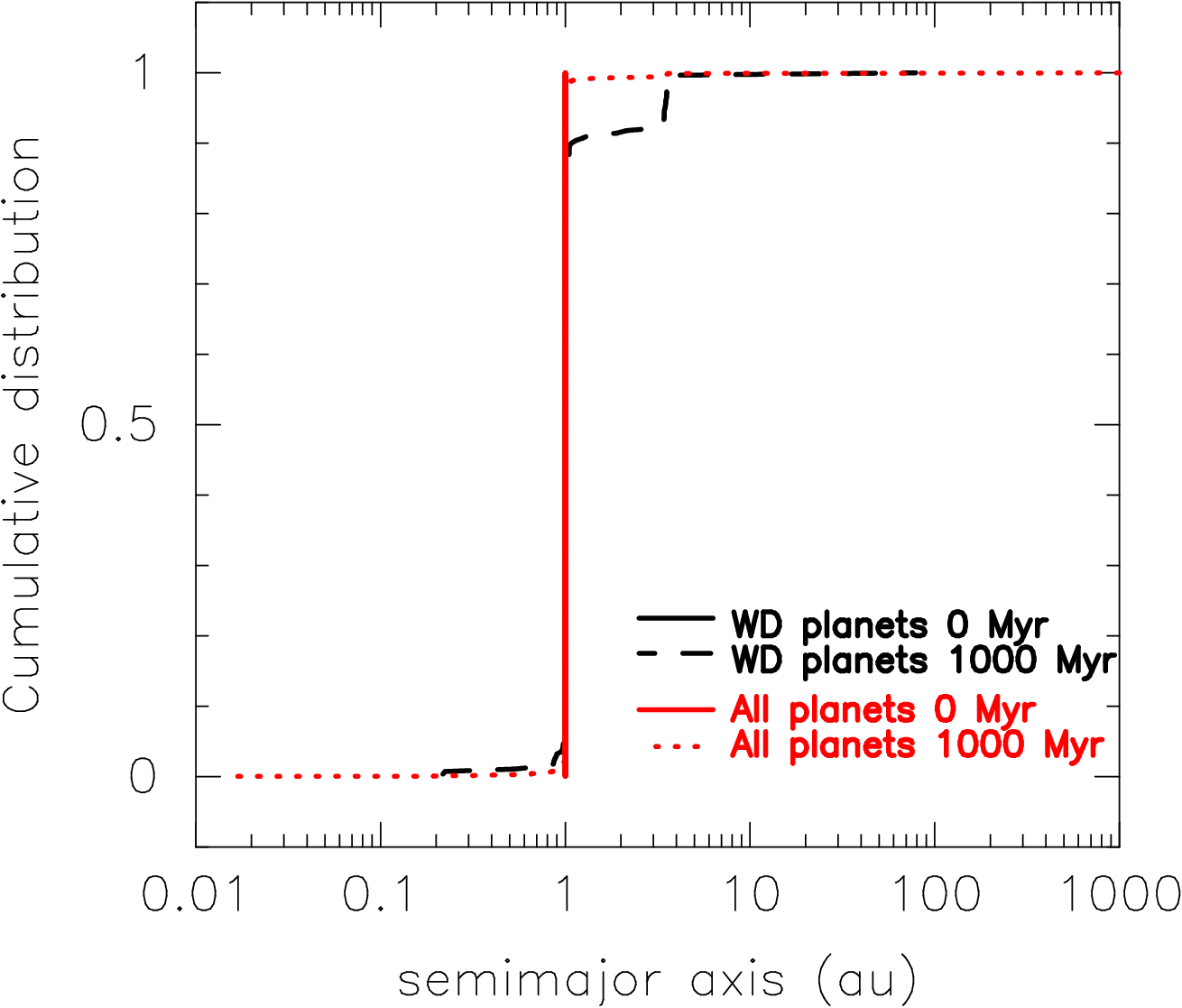}}}
\hspace*{0.3cm}\subfigure[30\,au]{\label{sma_0p1_au-c}\rotatebox{270}{\includegraphics[scale=0.25]{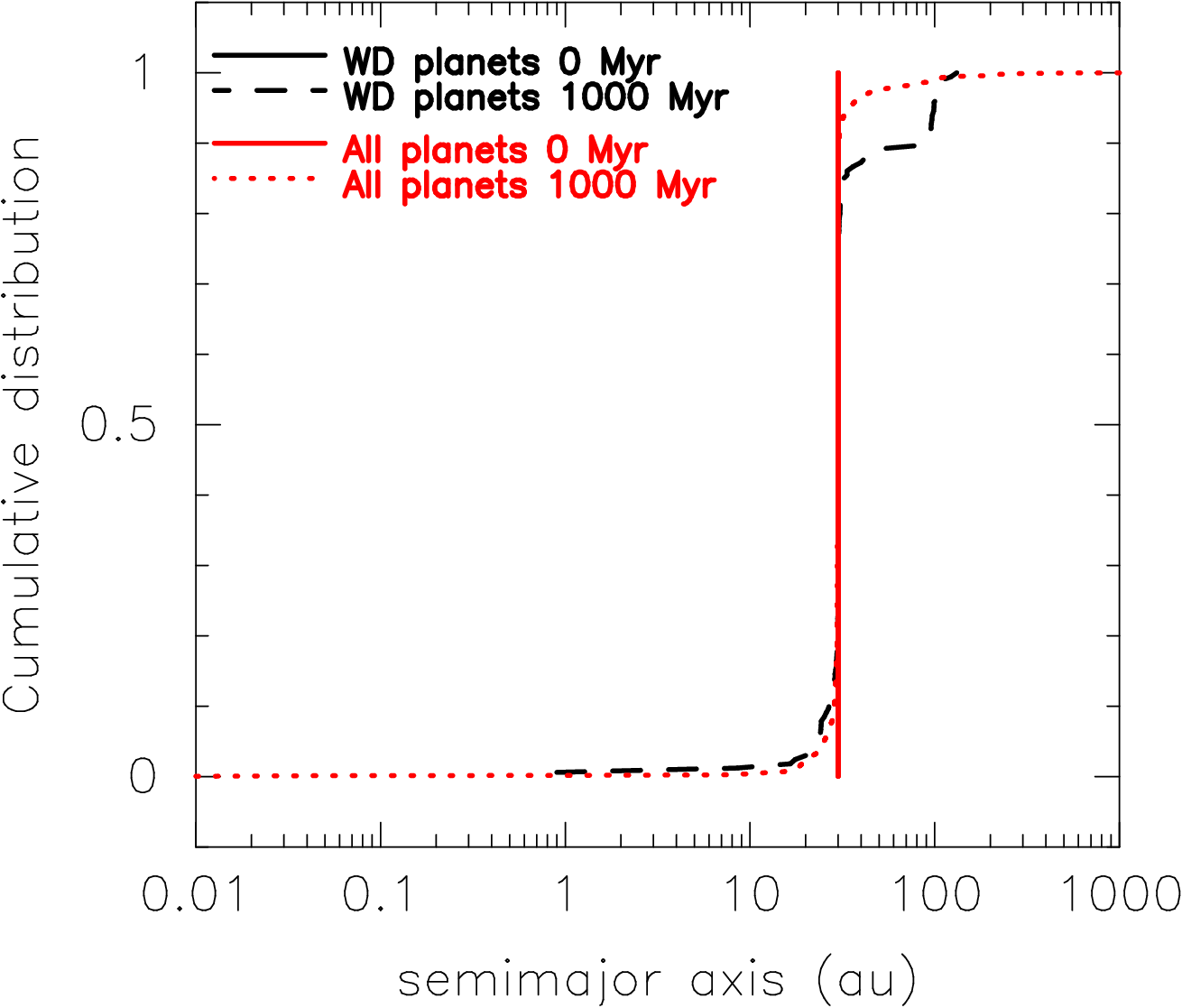}}}

\caption[bf]{Semimajor axes distributions of planets in dense ($\tilde{\rho} \sim 10^4$\,M$_\odot$\,pc$^{-3}$) star-forming regions where the planets shown are those that orbit their original (parent) star. The initial semimajor axes are drawn from delta functions at 0.1\,au (a), 1\,au (b) and 30\,au (c) -- simulation sets `C', `D' and `E', respectively. The distributions are taken at 1\,Gyr. The cumulative distributions of the semimajor axes of all planets are shown in red and white dwarf planets are shown in black.  
}
\label{sma_delta}
  \end{center}
\end{figure*}

In panel (a) we show the cumulative distributions of all planets (the red lines) and planets that formed around white dwarf progenitors (black lines). The histograms for the white dwarf planet semimajor axes are shown in panel (b) and the histograms for all planets are shown in panel (c). In both (b) and (c) the open histogram shows the initial distributions and the filled grey histogram shows the final distributions.

In Fig.~\ref{sma_uniform} there is a noticeable difference between the initial and final semimajor axes distributions of the white dwarf planets between around 40 -- 200\,au that is not apparent in the distributions of all planets, suggesting that the feature in the white dwarf planet distributions is not caused by dynamical encounters.

This is further corroborated if we examine the semimajor axes distribution before stellar evolution has occurred in any of the planet-hosting stars. In Fig.~\ref{sma_uniform_100myr} we show the same simulations but take the final snapshot after only 100\,Myr  (we choose 100\,Myr because this is before any planet-hosting stars have lost significant mass, but after the majority of dynamical encounters have occurred). Many of the planetary systems have experienced dynamical encounters, but the difference between the initial and final semimajor axis distributions for the white dwarf planets seen after 1\,Gyr is not there after 100\,Myr.

This happens because the planets orbiting white dwarfs experience an \emph{increase in their semimajor axis due to stellar evolution} which has more of an impact on the final distribution than dynamical encounters. Even in low-density star-forming regions ($r_F = 5$\,pc, $\tilde{\rho} = 10^2$\,M$_\odot$\,pc$^{-3}$ -- simulation set `B' in Table~\ref{cluster_sims}) where there are fewer dynamical encounters, the feature in the final white dwarf planet semimajor axis distribution is prominent. We show this in Fig.~\ref{sma_uniform_lowrho}, which is the initial and final semimajor axes distributions for planets in low-density star-forming regions. Fewer planetary systems are destroyed compared to the dense regions, and fewer planets have their orbits drastically altered by scattering dynamics. However, the increase in the semimajor axes of white dwarf planets after the stellar mass-loss is still apparent. 

The feature is largely independent of the assumed semimajor axis distribution (Fig.~\ref{sma_delta}), unless the semimajor axes are all very small $\ll 1$\,au, in which case the majority of the planets are engulfed before their orbits widen. We show this in Fig.~\ref{sma_delta}, where panel (a) is the semimajor axes distributions in simulations where all the planets are initially at 0.1\,au, panel (b) is where the planets are all initially at 1\,au, and panel (c) is where the planets are all initially at 30\,au.

When the planets are all at 0.1\,au, the binding energy is so high that the majority remain bound to the star and are engulfed before their semimajor major axis can increase due to stellar mass loss. When the planets are initially at 1 or 30\,au, some of the planets remain orbiting the white dwarfs, but at larger semimajor axes than their initial values, and this is again mainly due to mass-loss from the star, rather than any dynamical effects due to the cluster environment.

\subsubsection{Eccentricities}

\begin{figure}
\begin{center}
\rotatebox{270}{\includegraphics[scale=0.35]{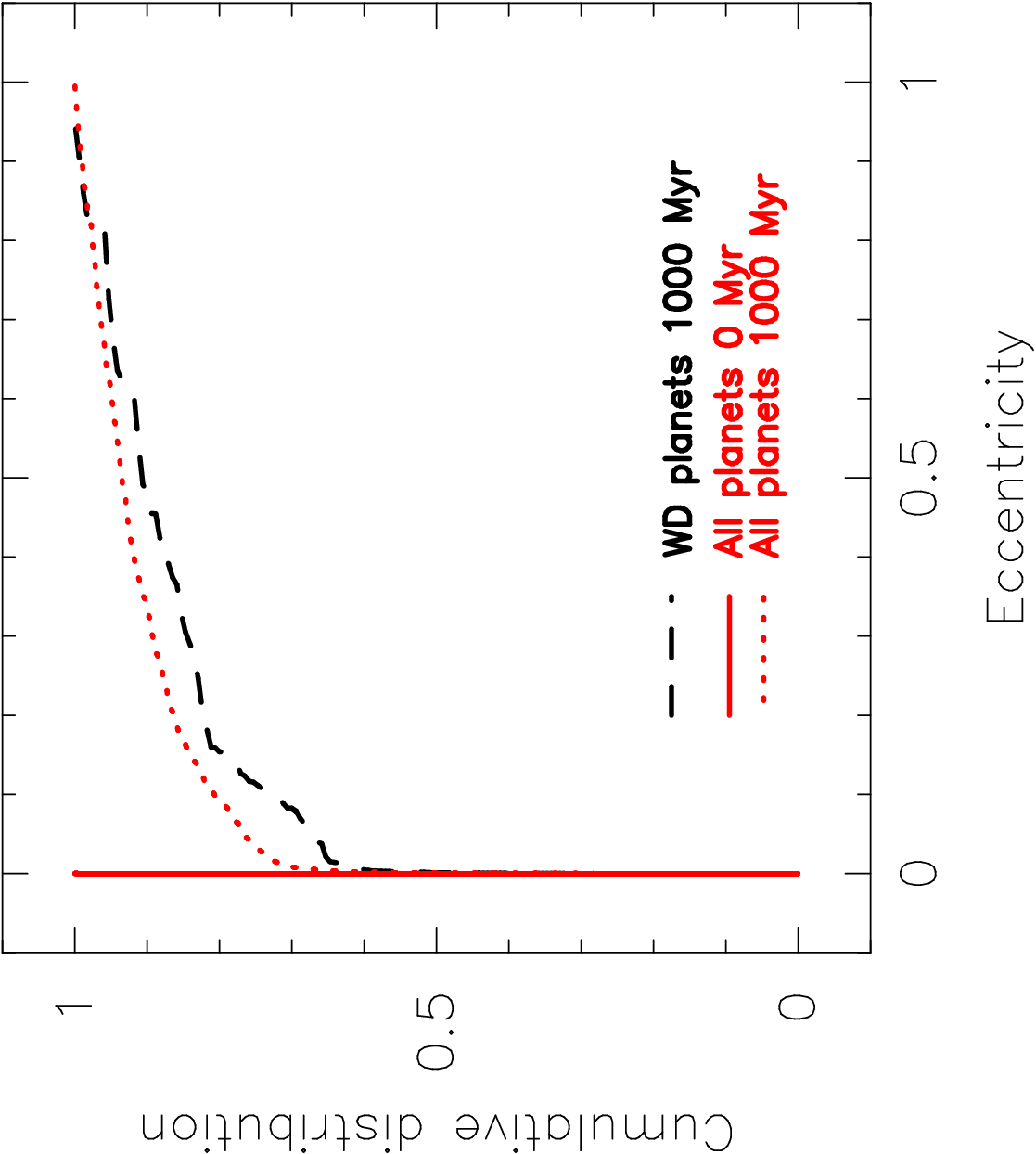}}
\end{center}
\caption[bf]{Eccentricity distributions in dense ($\tilde{\rho} \sim 10^4$\,M$_\odot$\,pc$^{-3}$) star-forming regions after 1\,Gyr of dynamical and stellar evolution. The black lines are white dwarf planets and the red lines are all planets. The planet semimajor axes are drawn from a uniform distribution between 0.1 -- 50\,au (simulation set `A'). We show the cumulative distributions of the semimajor axes of all planets (red) and white dwarf planets (black) -- the initial distributions are identical.  The planets shown are those that orbit their original (parent) star.}
\label{ecc_uniform}
\end{figure}

In Fig.~\ref{ecc_uniform} we show the eccentricity distributions of the white dwarf planets (black dashed line) and all planets (red dotted line) after 1\,Gyr in the simulations of dense ($\tilde{\rho} \sim 10^4$\,M$_\odot$\,pc$^{-3}$) star-forming regions where the planet semimajor axes are drawn from a flat distribution between 0.1 -- 50\,au (simulation set `A'), and the initial eccentricities are all set to zero. 40\,per cent of white dwarf planets have their eccentricities increased, compared to 30\,per cent of all planets. The change in eccentricity is almost entirely due to dynamical encounters \emph{before} the white dwarf phase, rather than after the star has become a white dwarf. The difference between the white dwarf planets and all planets is because higher mass stars typically experience more encounters, which in turn leads to more disruption of their planetary systems.

\subsubsection{Periastron distances}

\begin{figure*}
  \begin{center}
\setlength{\subfigcapskip}{10pt}
\hspace*{-1.5cm} \subfigure[]{\label{peri_uniform-a}\rotatebox{270}{\includegraphics[scale=0.35]{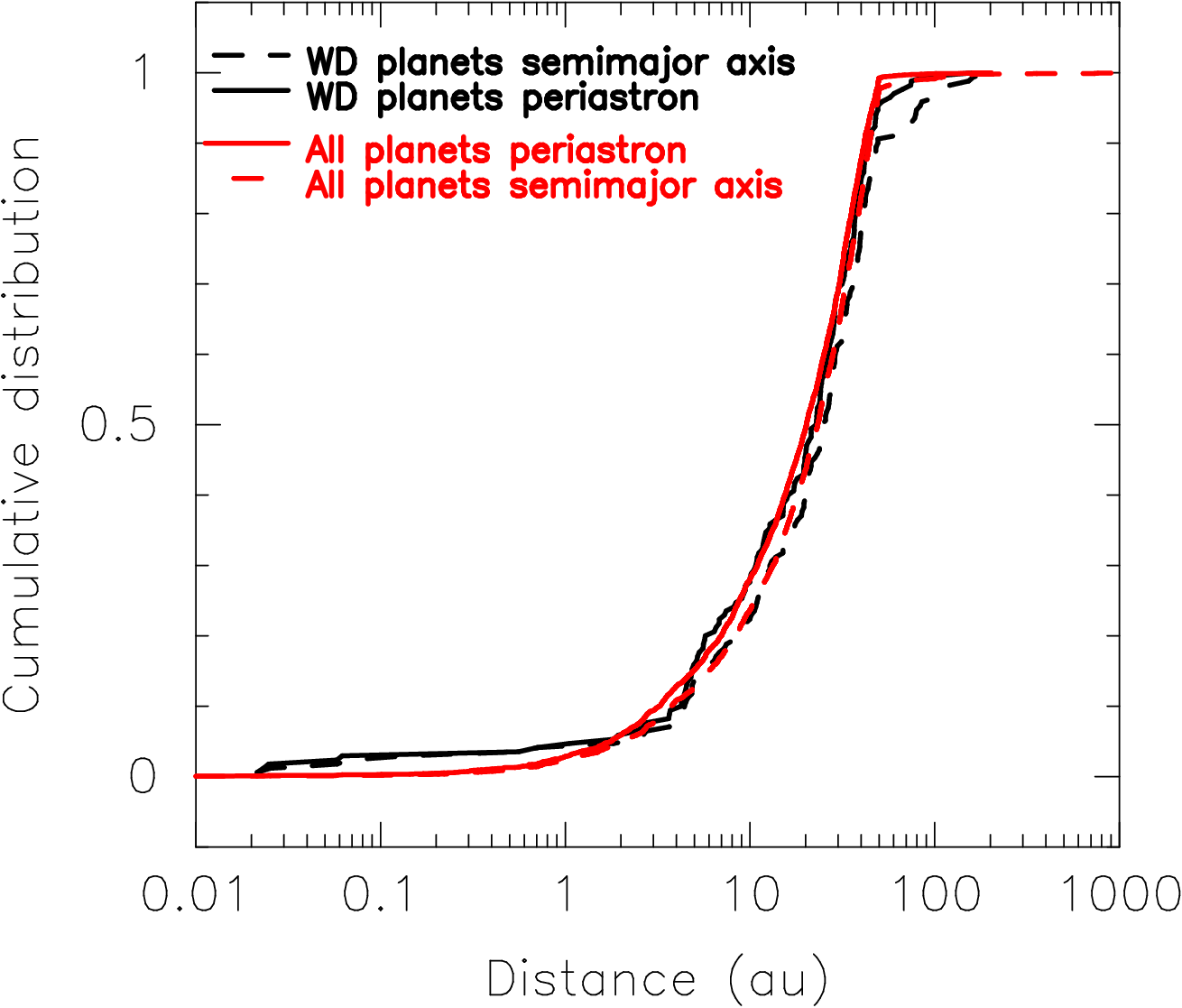}}} 
\hspace*{0.3cm}\subfigure[WD planets]{\label{peri_uniform-b}\rotatebox{270}{\includegraphics[scale=0.35]{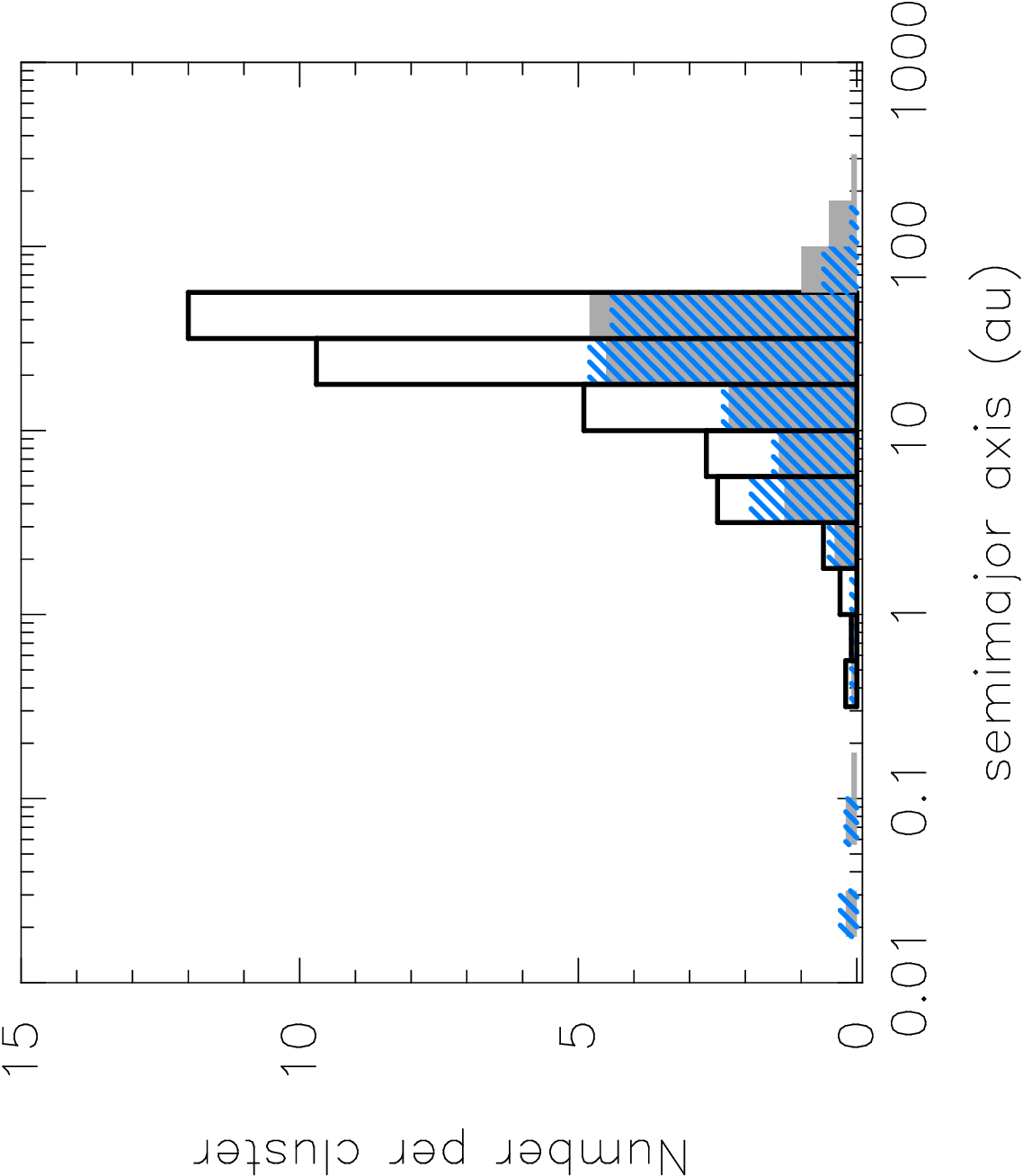}}}

\caption[bf]{The distributions of periastron distances for white dwarf planets. Panel (a) shows the cumulative distributions of the periastron distances (solid lines) compared to the semimajor axes (dashed lines) after 1\,Gyr in the default simulations of dense star-forming region where the initial planet semimajor axes are drawn from a uniform distribution between 0.1 -- 50\,au (simulation set `A').  Panel (b) shows the histogram for the initial (open) and final (grey fill) semimajor axes distributions for the white dwarf planets. The final periastron distances are shown by the blue hatched histogram.
}
\label{peri_uniform}
  \end{center}
\end{figure*}

Periastron distances of white dwarf planets represent particularly important parameters for both detectability prospects and because of their influence on whether their parent star will be polluted, and by how much, and how often. We now use the eccentricities with the semimajor axes to calculate the periastron distances of the planetary systems in our default simulation, where the planets' semimajor axes are drawn from a uniform distribution between 0.1 -- 50\,au. In Fig.~\ref{peri_uniform-a} we show the cumulative distributions of the final periastron distances by the solid lines and the final semimajor axes by the dashed lines (after 1\,Gyr). 

We also show the histogram of the white dwarf planets' periastron distances by the blue hatched histograms in Fig.~\ref{peri_uniform-b} superimposed on the histograms of the semimajor axes.

The periastron distance distributions tend to hide the difference in semimajor axes between the white dwarf planets and all planets caused by stellar mass-loss because overall the white dwarf planets have larger eccentricities and this skews the periastron distributions towards smaller values. It does not necessarily mean that the white dwarf  planets whose semimajor axes are increased due to stellar mass-loss have high eccentricities.

\subsection{Captured planets}

\begin{figure}
\begin{center}
\rotatebox{270}{\includegraphics[scale=0.35]{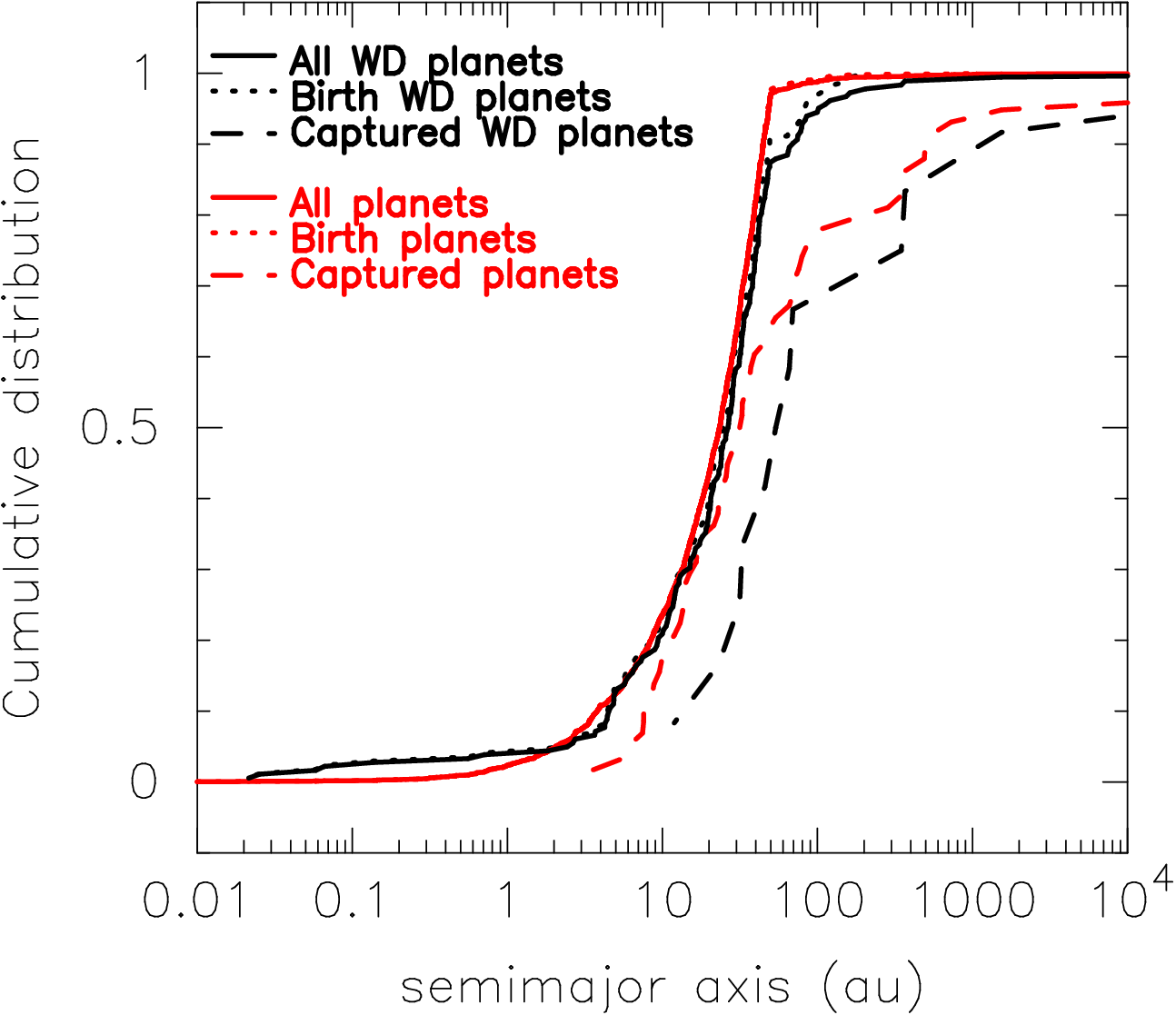}}
\end{center}
\caption[bf]{Semimajor axis distributions in our dense star-forming regions (simulation set `A') after 1\,Gyr. The black lines are white dwarf planets and the red lines are all planets. The distributions for the birth planets are shown by the dotted lines, and the distributions for the captured planets are shown by the dashed lines. The summed distributions (captured and birth systems) are shown by the solid lines.}
\label{sma_uniform_capt}
\end{figure}

A small number of planets form via capture when they form a bound orbit with another star that is different to the one they formed around, after being free-floating. The contribution to the semimajor axis distributions from captured planets in our default simulation set (`A') is shown in Fig.~\ref{sma_uniform_capt}. As before, white dwarf planets are shown by the black lines, and all planets are shown by the red lines. The lines on this plot are all the final distributions after 1\,Gyr; the dotted lines are the birth systems only, the dashed lines are the captured systems only, and the solid lines are all planetary systems. Whilst the contribution from captured planets does slightly alter the white dwarf planet semimajor axis distribution, the effects of stellar evolution still dominate over encounter dynamics for the white dwarf planets. 

\begin{figure}
\begin{center}
\rotatebox{270}{\includegraphics[scale=0.35]{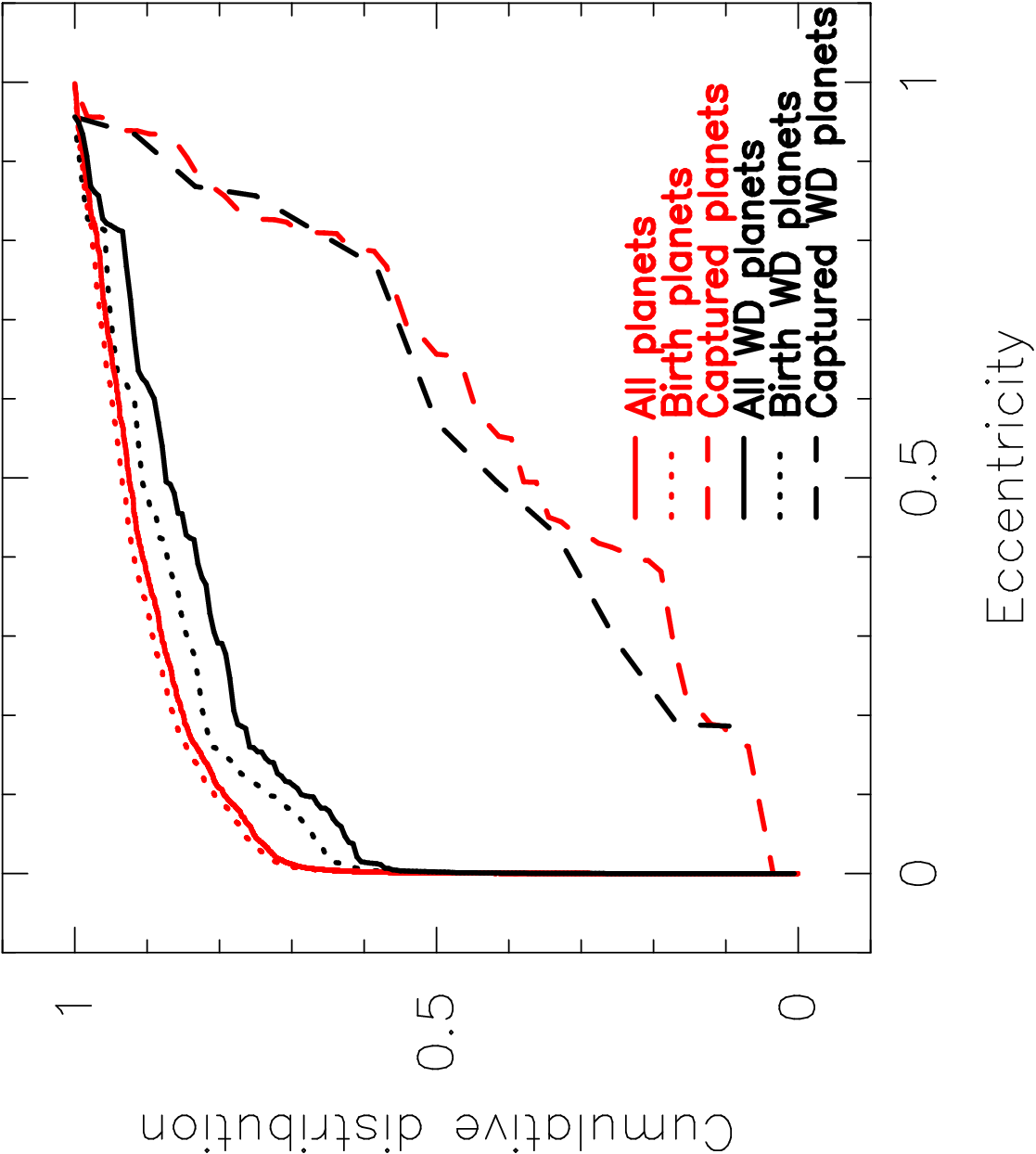}}
\end{center}
\caption[bf]{Eccentricity distributions in our dense star-forming regions (simulation set `A') after 1\,Gyr. The black lines are white dwarf planets and the red lines are all planets. The distributions for the birth planets are shown by the dotted lines, and the distributions for the captured planets are shown by the dashed lines. The summed distributions (captured and birth systems) are shown by the solid lines.}
\label{ecc_uniform_capt}
\end{figure}

The contribution to the eccentricity distributions from the captured planets is shown in Fig.~\ref{ecc_uniform_capt}. The lines on this plot are all the final distributions after 1\,Gyr; the dotted lines are the birth systems only, the dashed lines are the captured systems only, and the solid lines are all planetary systems. As with the semimajor axes distributions, the eccentricities of the captured systems do skew the distributions towards higher eccentricities, ostensibly because the captured systems tend to have a thermal ($f(e) \propto e^2$) distribution \citep{Kouwenhoven10,Perets12,DaffernPowell22}.

\subsection{White dwarf planets in triple systems}

The initial kinematic substructure in star-forming regions is conducive to the formation of multiple systems via capture, as nearby stars (and planets) will have similar velocities \citep{Kouwenhoven10,Moeckel10,Parker14d,DaffernPowell22,Parker22b}. In our simulations several triple systems form through this capture mechanism.

\begin{figure*}
  \begin{center}
\setlength{\subfigcapskip}{10pt}
\hspace*{-1.5cm} \subfigure[]{\label{bin_plots-a}\rotatebox{270}{\includegraphics[scale=0.25]{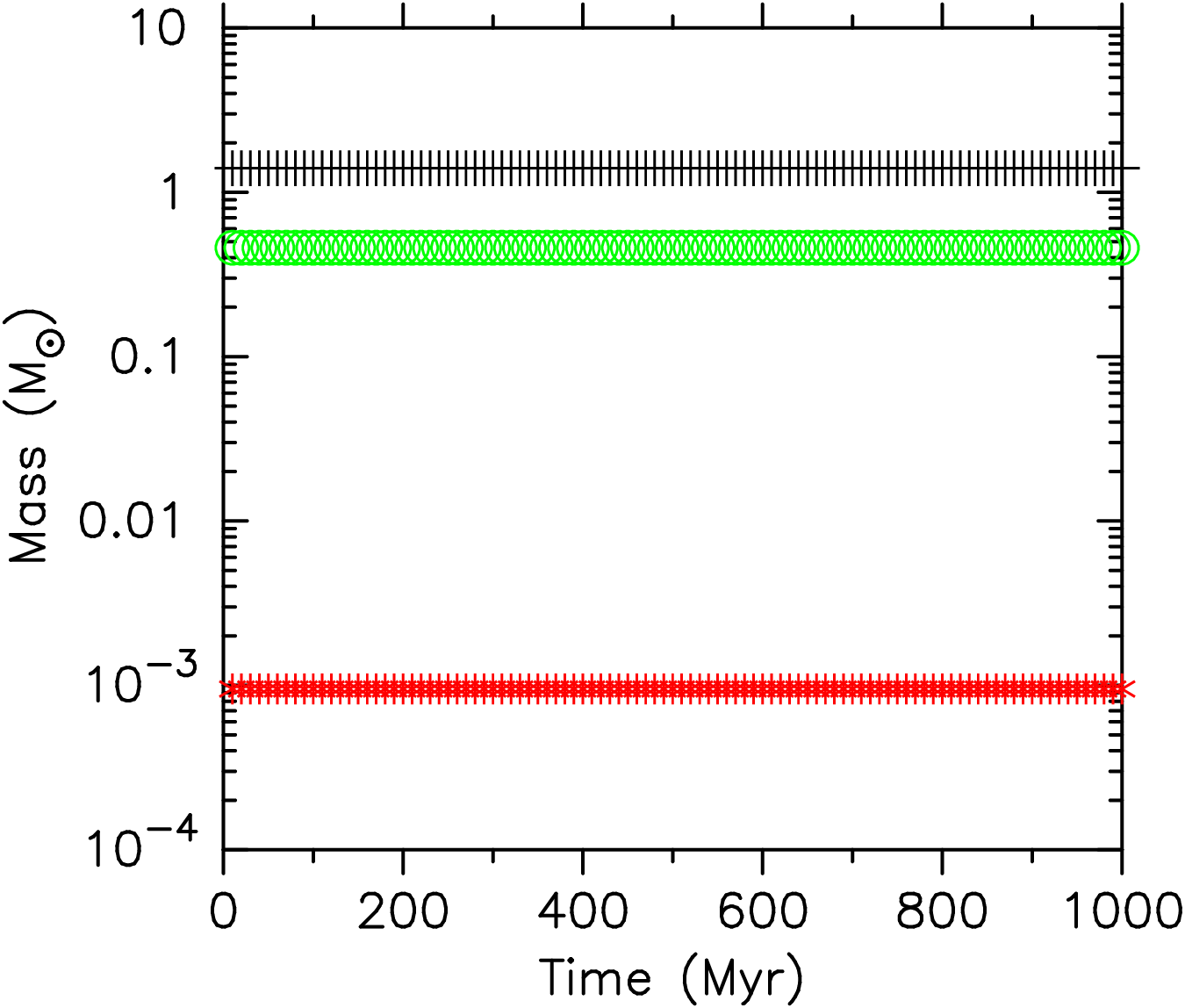}}} 
\hspace*{0.3cm}\subfigure[]{\label{bin_plots-b}\rotatebox{270}{\includegraphics[scale=0.25]{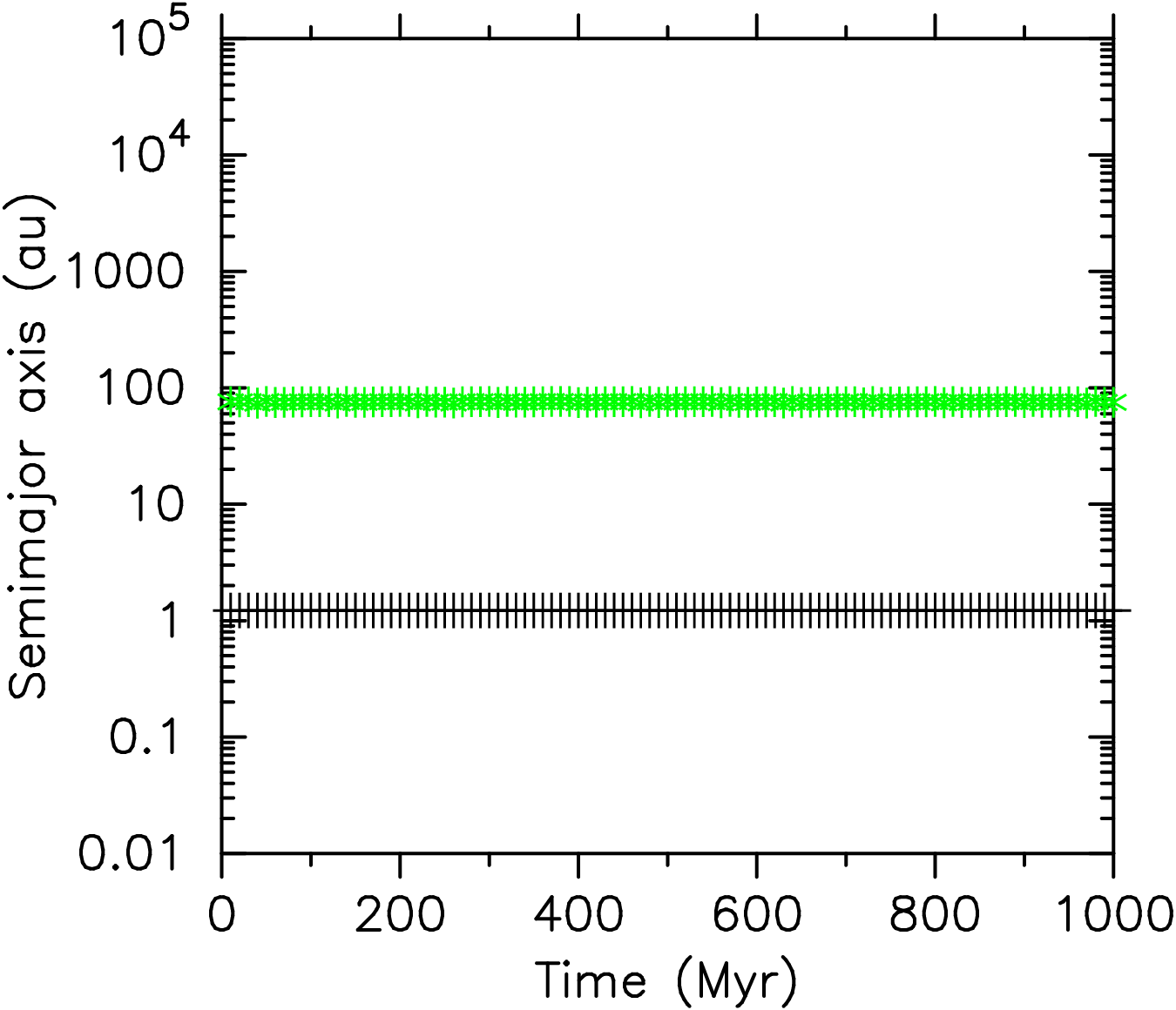}}}
\hspace*{0.3cm}\subfigure[]{\label{bin_plots-c}\rotatebox{270}{\includegraphics[scale=0.25]{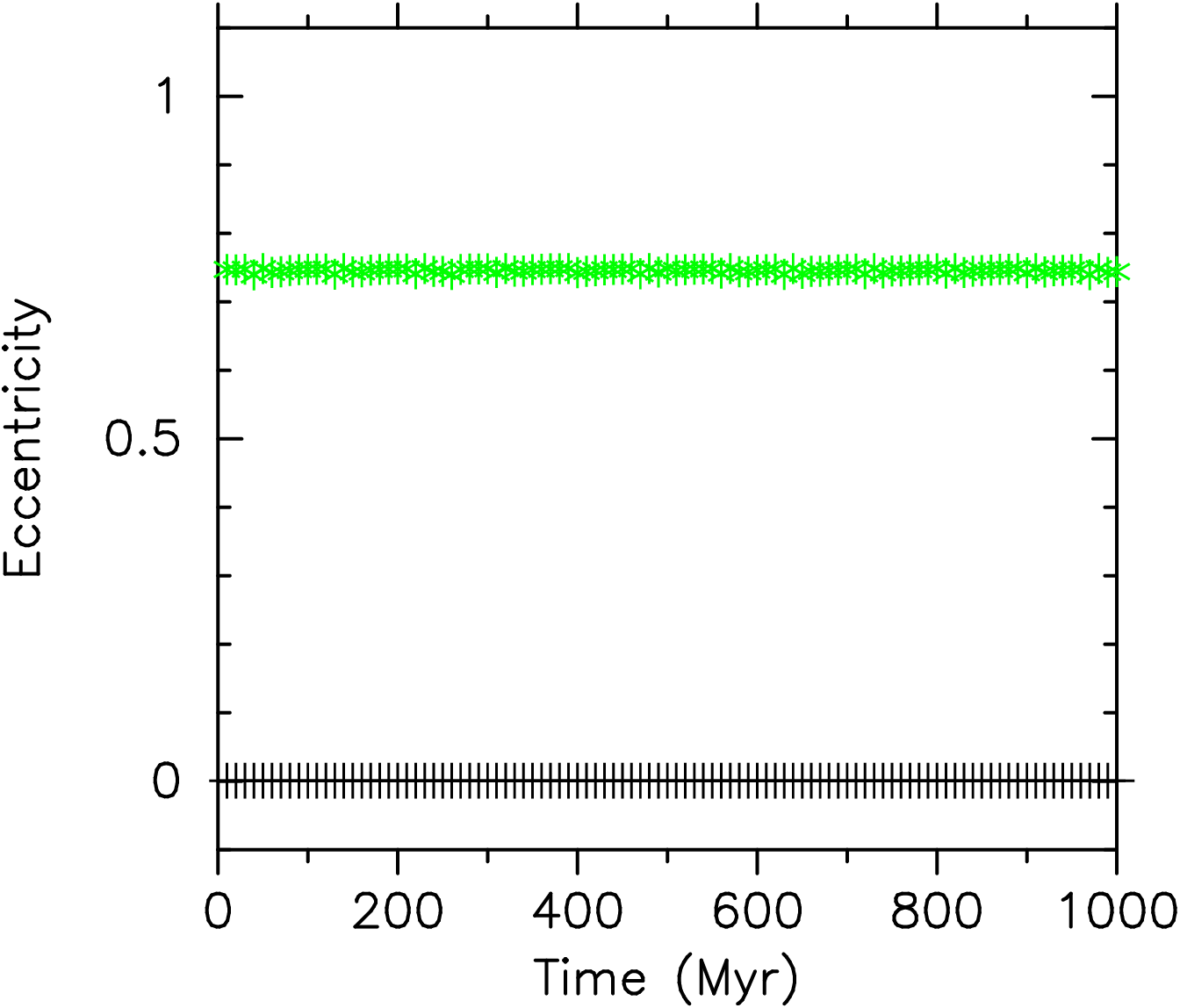}}}
\hspace*{-1.5cm} \subfigure[]{\label{bin_plots-d}\rotatebox{270}{\includegraphics[scale=0.25]{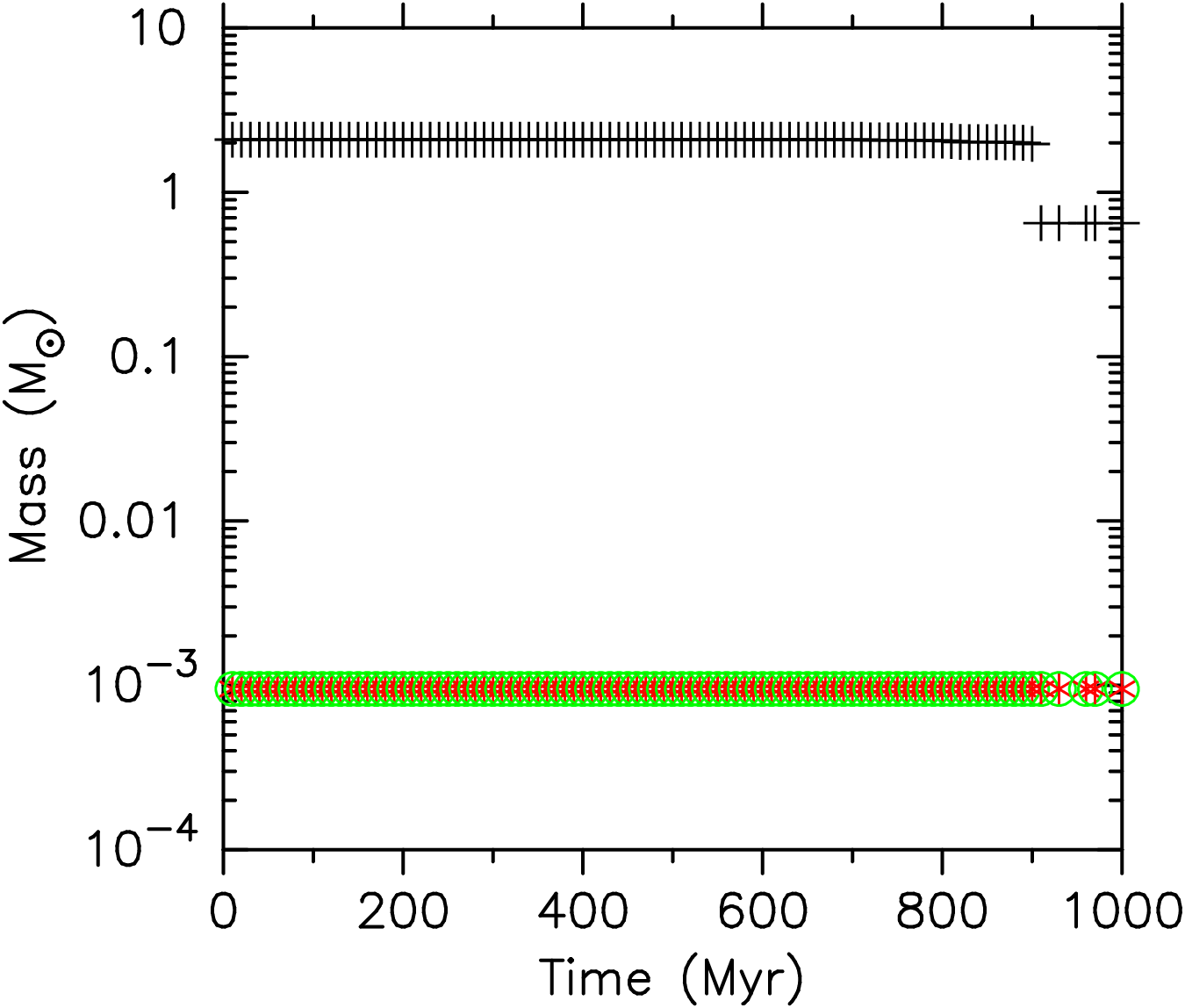}}} 
\hspace*{0.3cm}\subfigure[]{\label{bin_plots-e}\rotatebox{270}{\includegraphics[scale=0.25]{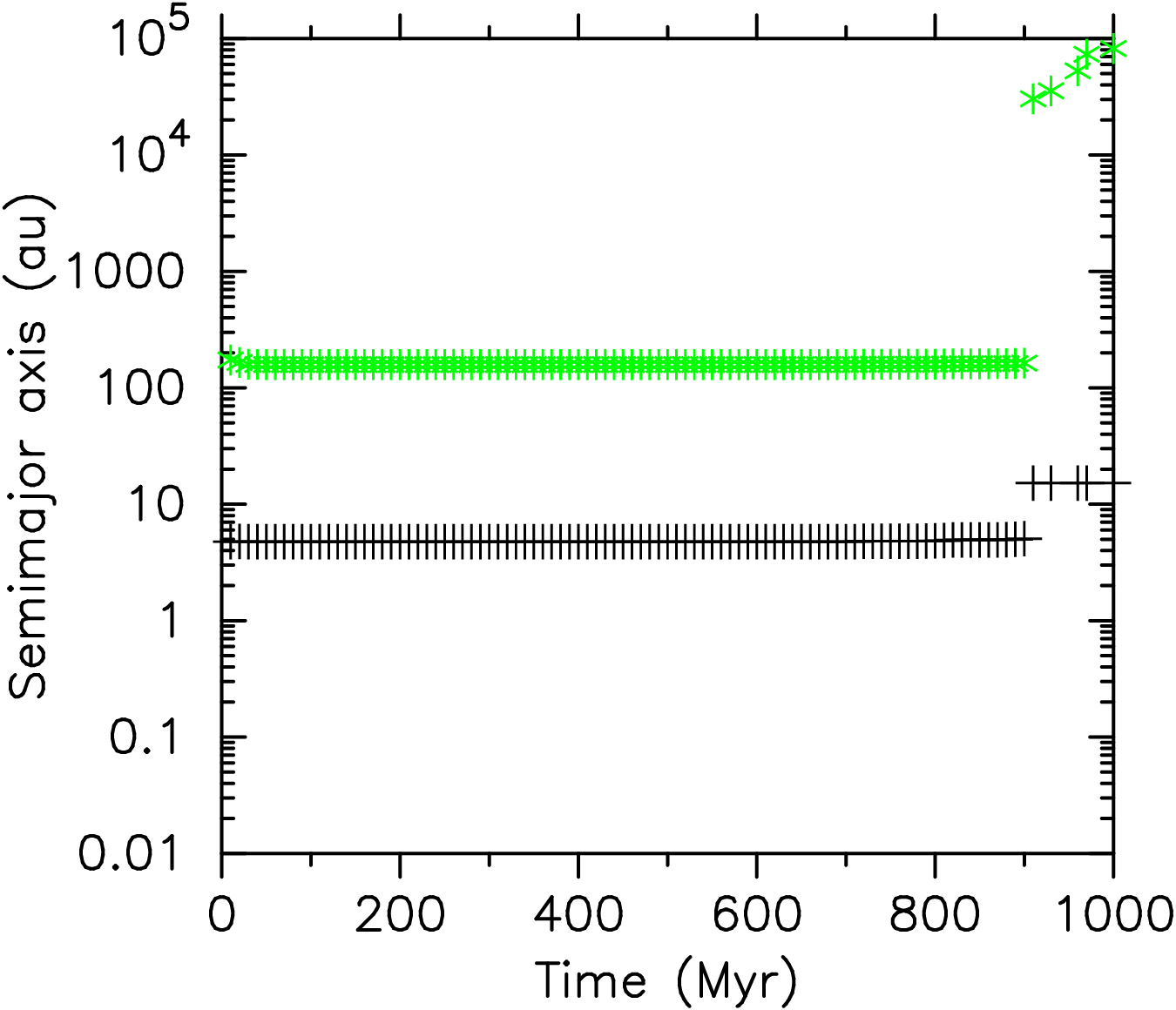}}}
\hspace*{0.3cm}\subfigure[]{\label{bin_plots-f}\rotatebox{270}{\includegraphics[scale=0.25]{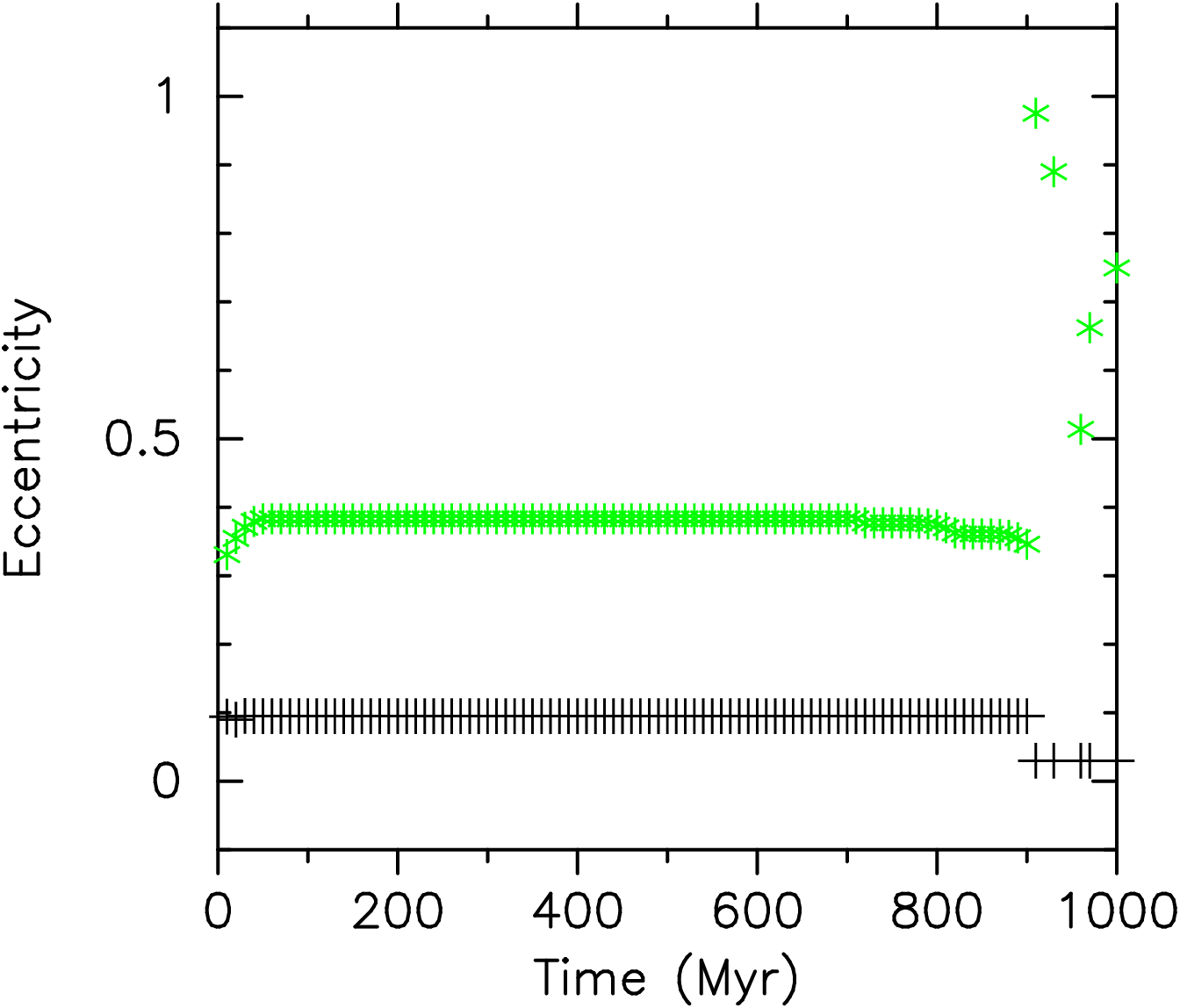}}}
\hspace*{-1.5cm} \subfigure[]{\label{bin_plots-g}\rotatebox{270}{\includegraphics[scale=0.25]{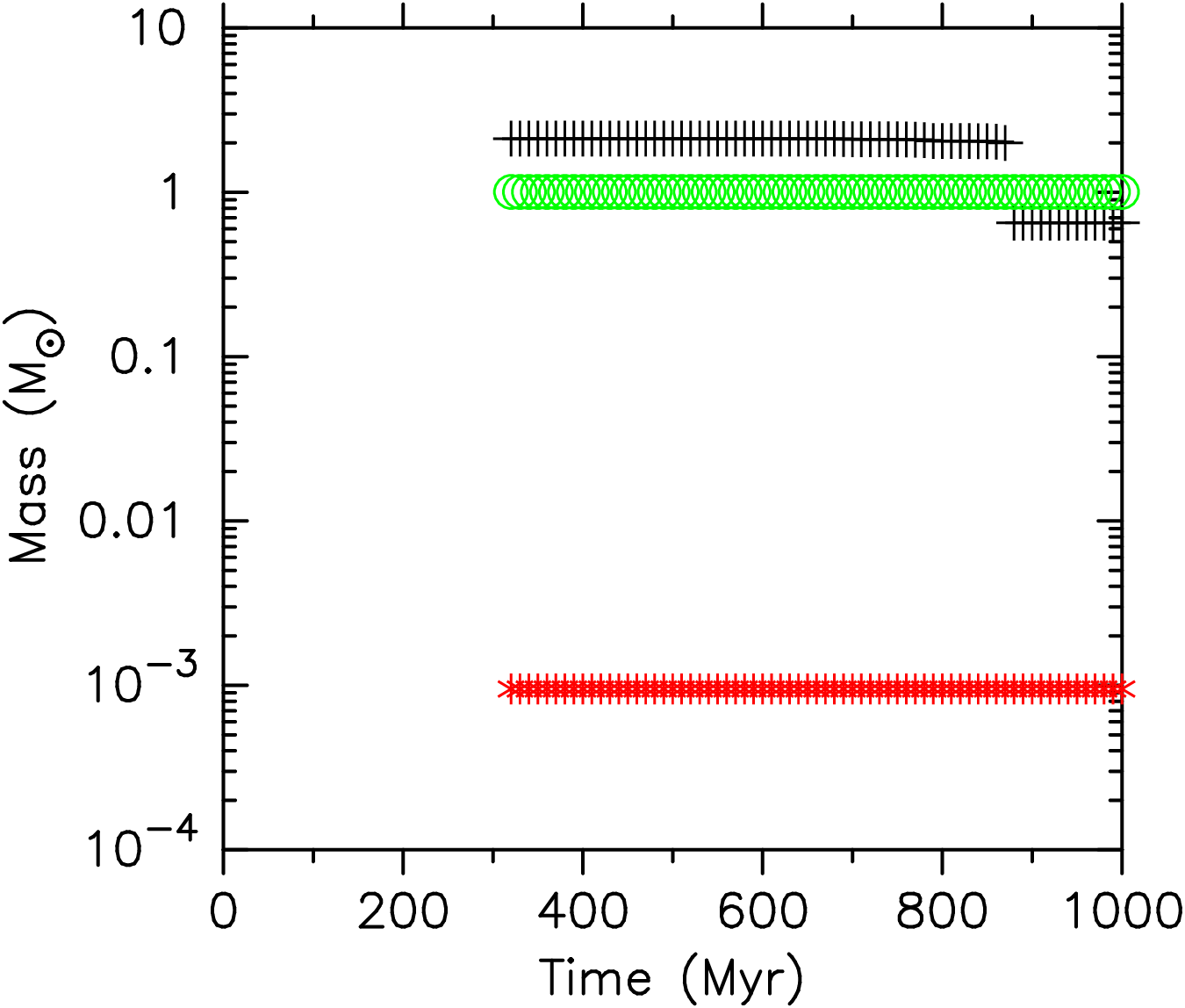}}} 
\hspace*{0.3cm}\subfigure[]{\label{bin_plots-h}\rotatebox{270}{\includegraphics[scale=0.25]{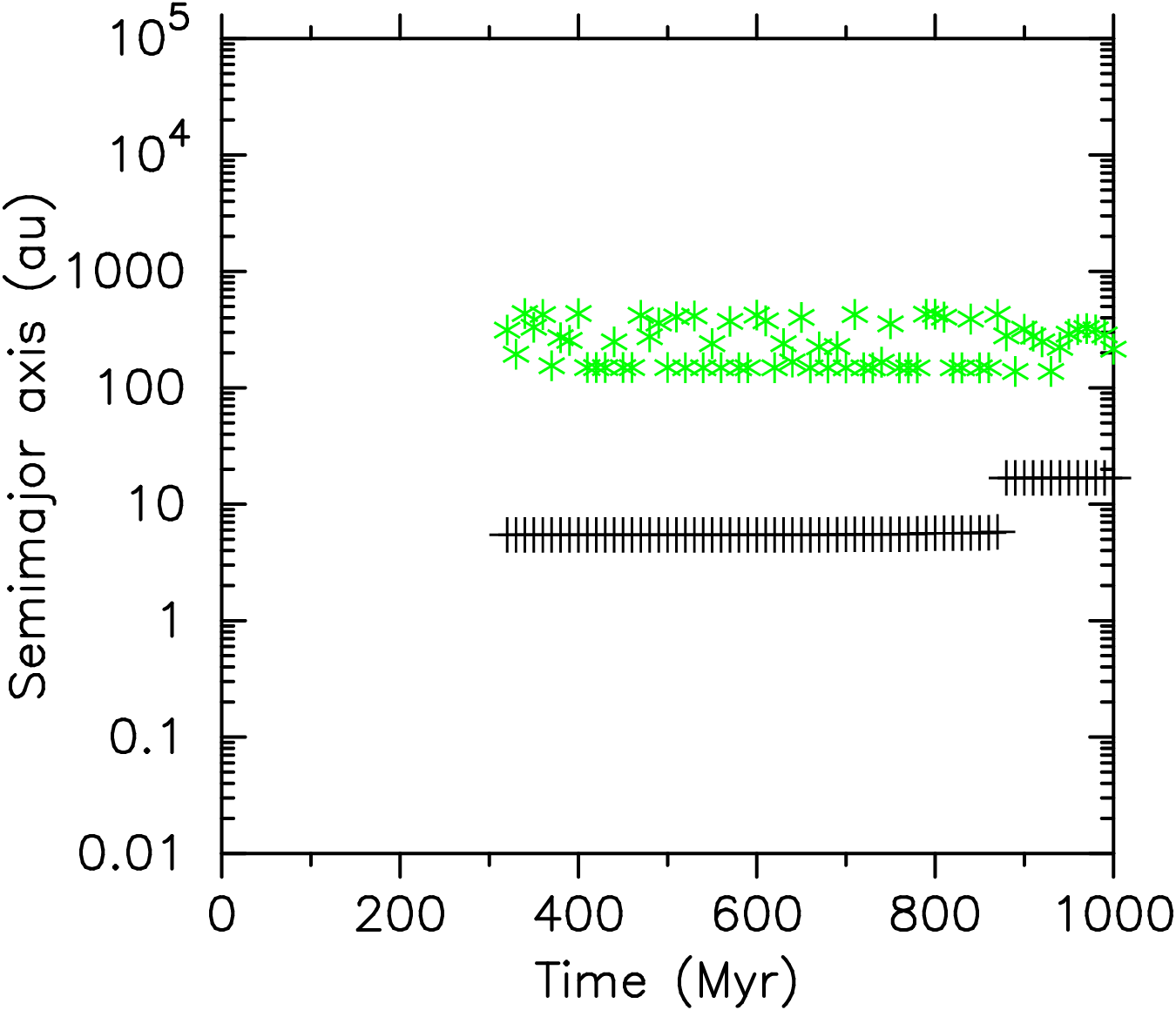}}}
\hspace*{0.3cm}\subfigure[]{\label{bin_plots-i}\rotatebox{270}{\includegraphics[scale=0.25]{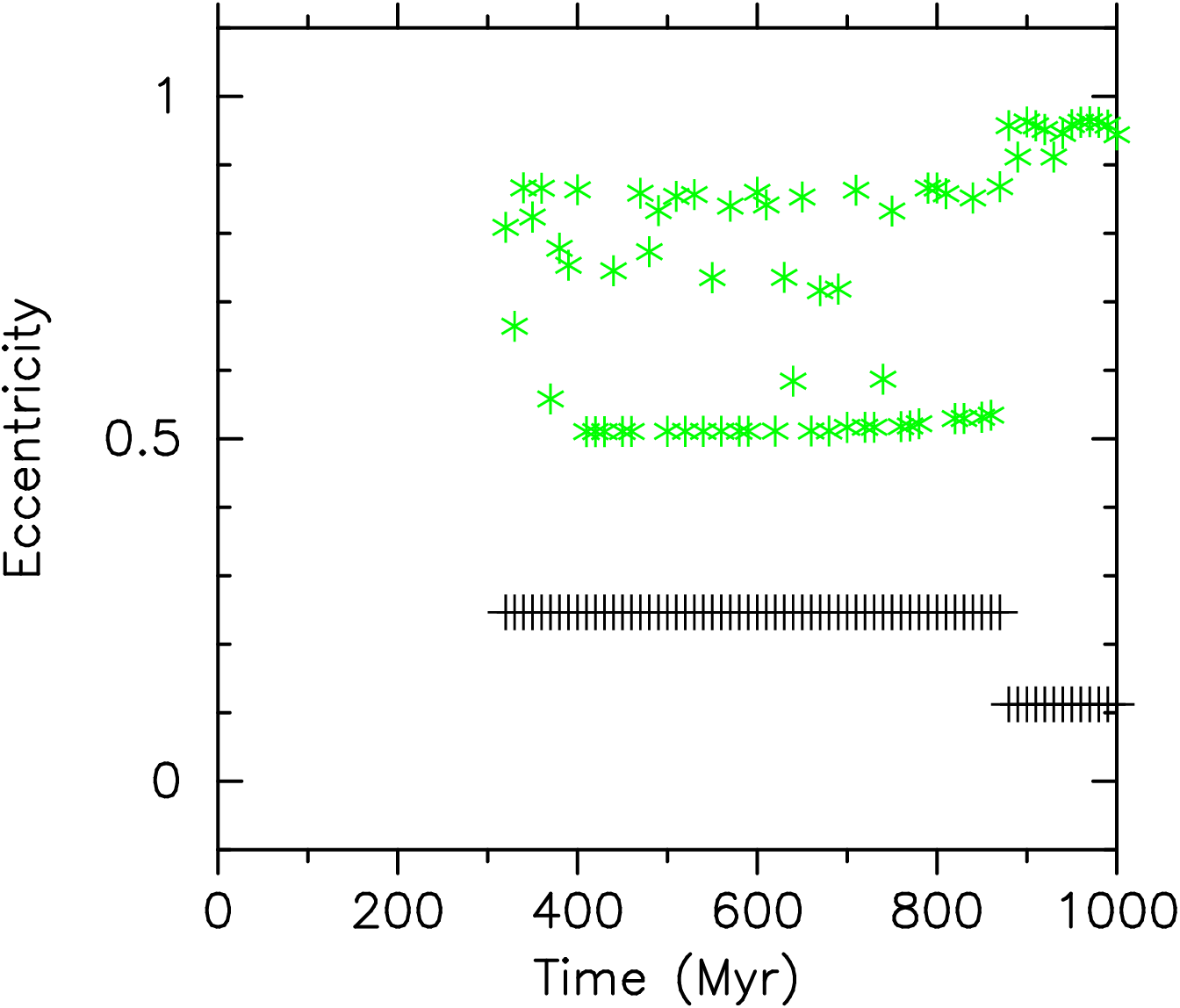}}}

\caption[bf]{Three examples of planetary systems in our default simulations (set `A') where the host star is more massive than 1\,M$_\odot$ and the system forms a triple with either another star, or a free-floating planet. The lefthand panels (a,d,g) show the evolution (if any) of the mass with time, with the planet-host star shown in black, the planet shown in red, and the captured object shown in green. The middle panels (b,e,h) show the evolution of the semimajor axes of the subsystems, with the black symbols showing the inner semimajor axis, and the green points showing the outer semimajor axis. The righthand panels (c,f,i) show the evolution of the eccentricity of these systems, with the black points showing the evolution of the eccentricity of the inner system, and the green symbols showing the evolution of the eccentricity of the outer system. The top row (a,b,c) shows a system with a 1.4\,M$_\odot$ star that remains on the main sequence, which in the first 10\,Myr forms a triple system with a 0.46\,M$_\odot$ star.  The second row (d,e,f) shows a system with a 2.09\,M$_\odot$ star that evolves to a white dwarf after 910\,Myr. The system captures a free-floating planet within the first 10\,Myr. In the bottom panels (g,h,i) we show a system that first forms after 320\,Myr and consists of a 2.12\,M$_\odot$ star that becomes a white dwarf after 880\,Myr. The outer star is a 1.00\,M$_\odot$ star that remains on the main sequence throughout the simulation. 
}
\label{bin_plots}
  \end{center}
\end{figure*}

Over the ten realisations of our default simulation initial conditions (set `A'), 11 longlived (>500\,Myr) triple systems form (i.e. $\sim$1 per cluster) containing a star-planet where the star will eventually evolve into a white dwarf. In some of these systems the host star becomes a white dwarf during the simulation 1\,Gyr run time. The captured object is usually another star, but in three out of eleven systems it is a planet that had been free-floating in the cluster.

In Fig.~\ref{bin_plots} we show three systems that have formed a triple with a star--planet as the central binary, with a third captured object. For each system we show the evolution with time of (i) the masses of the three components (lefthand panels, where the planet-hosting star is shown by the black symbols, the planet is shown by the red symbols, and the captured object is shown by the green symbols); (ii) the semimajor axes (middle panels) and (iii) eccentricity (right panels). The semimajor axes and eccentricities of the inner (star-planet) system is shown in black, and the outer semimajor axes and eccentricities are shown in green.

In the first triple system (top row) the planet host star is a 1.4\,M$_\odot$ star that does not leave the main sequence, and in the first 10\,Myr forms a triple system with a 0.46\,M$_\odot$ star. The semimajor axes and eccentricities remain constant.

In the second (middle row), we show a system with a 2.09\,M$_\odot$ star that evolves to a white dwarf after 910\,Myr. Initially, its planet is at 4.8\,au and this orbit expands to 15.2\,au in the white dwarf phase. The system captures a free-floating planet within the first 10\,Myr on an orbit of 160\,au and eccentricity that varies between 0.3 and 0.4 during the main sequence phase of the host star. When the star becomes a white dwarf the outer planet's semimajor axis increases to above $10^4$\,au and the eccentricity fluctuates between 0.6 and 0.97 in the white dwarf phase. The eccentricity of the inner planet decreases during the white dwarf phase.

In the final system we show (bottom row), the capture occurs long after most of the dynamical evolution in the cluster has occurred (320\,Myr), and the system captures a 1\,M$_\odot$ star. The system consists of a 2.12\,M$_\odot$ star that becomes a white dwarf after 880\,Myr. The outer star is a 1.00\,M$_\odot$ star that remains on the main sequence throughout the simulation. The orbit of the planet increases from 5.8 to 16.8\,au during the white dwarf phase, but its eccentricity decreases from 0.25 to 0.11. The outer star's semimajor axis varies between 138\,au and 434\,au but does not dramatically increase during the white dwarf phase. The eccentricity  does increase during the white dwarf phase, approaching unity towards the end of the simulation.

\section{Discussion}
\label{discussion}

We have shown that even in environments where a high degree of dynamical processing between different planetary systems occurs, the semimajor axes distributions of these planets are clearly dominated by the mass-loss from the star during the post-main sequence phase. The reason for this is that even in a dense star-forming region, only $\sim$10 -- 20\,per cent of stars have their semimajor axes significantly changed by close encounters with stellar flybys \citep{Parker12a,DaffernPowell22}, whereas all planets that are not engulfed by the star will see their semimajor axes increase due to mass-loss from the post-main sequence evolution of the star.

\subsection{Observational implications}

Besides this main conclusion, our other result is more generally an approximate population synthesis of giant planets orbiting white dwarfs that are emerging from a birth cluster\footnote{\cite{MauchSoriano26} also modelled a population synthesis of giant planets around white dwarfs with an initial mass function, but they did not model birth cluster evolution. \cite{Veras20a} and \cite{Stock22} modelled birth cluster dynamics, but not with stellar evolution, and focused on other aspects of white dwarf planetary systems aside from giant planets.  The model of \citet{Veras24b} generates only minor planets, with a link to white dwarf planetary systems. Other population synthesis models for exoplanetary systems do not prominently feature white dwarfs, e.g. \citet{Ida04,Ida13,Schlecker21,Burn22,Chen25}.}. The resulting orbital parameter distributions have implications for our currently limited sample of known or strongly suspected white dwarf planets.

First consider WD~0806-661~b \citep{Luhman11}, a substellar-mass object whose mass is now constrained to be in the planetary regime \citep{Voyer25}. This planet and its host white dwarf are separated by 2500\,au on the sky; although we do not know its semimajor axis or eccentricity, this wide separation is potentially suggestive of a captured planet.

A small fraction (typically about 10\,per cent) of the white dwarf planetary systems in our simulations capture their planet. These systems typically have semimajor axes in the range 100s to 1000s\,au and high eccentricities. Hence, this outcome may be indicative of the dynamical origin of WD~0806-661~b. Planetary capture in a stellar cluster is much more likely than in the Galactic field because of the higher density in the cluster and the slower flyby speeds.

Next consider the planet WD~1856+534~b, which is the only known transiting planet around a white dwarf. This planet transits at a distance of $\approx$~0.02~{\rm au} \citep{Vanderburg20}. Because this distance is well within the typical engulfment distance for giant planets of a few au, several investigations have investigated this planet's dynamical origin. Proposed origins include second-generation formation \citep{Chamandy25}, common-envelope evolution \citep{Chamandy21,Lagos21,Merlov21} and gravitational scattering \citep{Munoz20,Vanderburg20,OConnor21,Stephan21}.

Arguably scattering, followed by tidal truncation, is the most likely origin, because the planet is unlikely to be massive enough to eject the stellar evolution \citep{OConnor23b}. However, the flavour of scattering is unconstrained; it could be due to another hidden giant planet in the system, one of the companion stars, or a stellar flyby. In our default simulations, three planetary systems have final semimajor axes that are between 0.01 -- 0.1\,au, and all of these systems had planets with initial semimajor axes between 1 -- 20\,au. The decrease in semimajor axes of these planets occurred during the star cluster phase due to scattering from flybys. Also, for similar systems which may be found in the future \citep{Robert24}, our results will help quantify the probability that this scattering could have occurred while within the stellar cluster.

A different planet with a similarly tight orbit (with $a\sim 0.07$~au) is WD~J0914+1914~b \citep{Gaensicke19}. In this system, the cooling age of the white dwarf is extremely young, at 13.3~Myr. Hence, gravitational scattering  and subsequent orbital shrinkage due to tides must have occurred quickly \citep{Veras20c}. Cluster flyby scattering, as modelled here, might then help leverage planet-planet scattering as a more- or less-likely possibility based on one's assumptions about the formation of multi-planet systems with multiple giant planets.

Now consider the first planet discovered orbiting a white dwarf, PSR~B1620-26~(AB)~b, which is a circumbinary planet orbiting a white dwarf and a pulsar at a distance of about 23\,au \citep{Thorsett93, Sigurdsson03}. Although our simulations do not model pulsars, our simulations do yield triple systems containing a planet and a white dwarf. That result alone demonstrates how such configurations may be obtained just from scattering within stellar clusters.

Finally, consider circumstellar planets orbiting white dwarfs in the 1 -- 100\,au range. At least three detection techniques can probe this range: microlensing, astrometry and direct imaging. Microlensing is currently the most successful white dwarf planet hunting technique, with two discoveries \citep{Blackman21,Zhang24}, and potentially many more to come with the Nancy Grace Roman Space Telescope. Astrometry should yield at least six planets with the final Gaia data release \citep{Sanderson22}, and direct imaging with JWST is close to succeeding \citep{Mullally24,Mullally26}. The results of our simulations help make predictions for all of these techniques given that the vast majority of our final planet semimajor axes lie within the range 1 -- 100\,au.

\subsection{Caveats and assumptions}

Clearly, our planetary `systems' are not systems in the sense that we do not (and cannot, due to technical limitations) model a full planetary system, including terrestrial planets and debris discs, in the simulations. In the future, it may be possible to include multi-planetary systems within a self-consistent $N$-body simulation, but current approaches can also evolve the planetary system between timesteps in the simulation of the star-forming region/cluster \citep{Cai17a,VanElteren19}.

However, our simulations are the first to model Jovian-mass planets in a clustered environment \emph{and} include stellar evolution via look-up tables. The stellar evolution code accessed by the $N$-body integrator is \texttt{SeBa} \citep{Zwart96,Zwart12,Toonen12}, which follows the main stages of stellar evolution, but other authors have shown that there are differences in the evolution of post-main sequence planetary systems that subtly depend on the adopted stellar evolution code \citep{MauchSoriano26}. Regardless of the adopted stellar evolution code, the significant mass-loss that occurs during the post-main sequence phase will still cause the orbits of bound planets to widen, and will therefore still change the semimajor axis distribution of the planets long after any dynamical evolution in the birth star-forming region.    

We have limited our primordial planetary systems to stars with initial masses between 0.1 -- 2.5\,M$_\odot$ although planetary systems have been observed around higher-mass stars \citep[e.g.][]{Janson21b,Squicciarini22,Delorme24}. However, high-mass progenitors of white dwarfs become increasingly rare as the mass increases \citep{OuldRouis24,Cunningham25} such that any progenitor masses above 5\,M$_\odot$ represent special cases \citep{Cheng25}.  
Nevertheless, we performed a test simulation where planets were placed around stars up to 5\,M$_\odot$. We found no significant differences between those runs and the simulations presented here.

The initial conditions we adopt for the star-forming regions may not be representative of the star-forming regions where the majority of exoplanets -- or white dwarfs -- formed. For example, our default simulations are of star-forming regions with a high degree of spatial and kinematic substructure \citep[which are informed by observations, e.g.][]{Larson81,Sanchez09,Hacar16}. This substructure is erased within a few Myr, and the exact degree of substructure has very little impact on the much longer-term evolution of these simulations \citep{Parker14b,Parker16c}. However, if a star-forming region has no kinematic structure \citep[e.g][]{Plummer11,Aarseth74}, then systems formed via capture are rare \citep{Kouwenhoven10,DaffernPowell22} and this would have implications for some of our results (though not our main result, which is the change in semimajor axes for planets that survive the post-main sequence evolution of their parent star).

Interestingly, the initial stellar density has no impact on our main result because the semimajor axis increase due to mass-loss is independent of the amount of dynamical evolution in the star-forming region/cluster. Similarly, other dynamical effects, such as the sudden disruption of the star-forming region due to gas removal \citep[e.g.][]{Tutukov78,Lada84,Goodwin97a,Shukirgaliyev18}, will not impact the results.

On a related point, there are uncertainties surrounding the timescales on which star-forming regions disperse into the Galactic field, but the majority of star-forming regions are thought to disperse long after 10\,Myr \citep[e.g.][]{Lada03}, whereas the majority of orbit-altering dynamical encounters occur within the first 10\,Myr of our simulations' evolution, regardless of the initial stellar density. The point at which the star begins to lose mass and could further alter the planet's orbit is therefore likely to occur after the star has left its birth environment. 

\section{Conclusions}
\label{conclusion}

We perform $N$-body simulations of star-forming regions with stellar evolution to determine the fate of Jovian-mass planets that orbit white dwarf progenitor stars ($1 \leq M_{\star}/{\rm M_\odot} < 2.5$). We follow the dynamical evolution in the star-forming region, as well as the individual evolution of the star-planet systems, once the star loses mass after it leaves the main sequence. Our conclusions are the following:

(i) Depending on the initial stellar density, a significant number of planetary systems are destroyed (up to 50\,per cent), and these planets either remain free-floating or are captured onto orbits around other single stars, or other star-planet systems.

(ii) Of the systems that remain bound, the expected change to their semimajor axis distribution from post-main sequence mass-loss is significantly modified by dynamical encounters in the system's birth environment in only about 20\,per cent of cases.

(iii) Hence, in the other 80\,per cent of cases, any currently observed giant planets orbiting white dwarfs within a few au must have arrived there through means other than birth cluster flyby scattering.

(iv) Around 10\,per cent of planets around white dwarf host stars are not primordial systems -- they form via capture of free-floating planets, and tend to have high semimajor axes ($a>$100\,au) and high eccentricities ($e>0.5$).\\

Our simulations therefore indicate that the effects of the star-forming environment on the properties of white dwarf planetary systems are modest, helping to constrain targeted searches for white dwarf planets. 






  

\section*{Data availability statement}

The data used to produce the plots in this paper will be shared on reasonable request to the corresponding author.

\section*{Acknowledgments}

 We thank the anonymous referee for a helpful report. RJP acknowledges support from the Royal Society in the form of a Dorothy Hodgkin Fellowship.

\bibliographystyle{mnras}
\bibliography{general_ref}

\appendix

\section{Different cluster initial conditions}

\begin{figure*}
  \begin{center}
\setlength{\subfigcapskip}{10pt}
\hspace*{-1.5cm} \subfigure[Fractal]{\label{density_evolution-a}\rotatebox{270}{\includegraphics[scale=0.35]{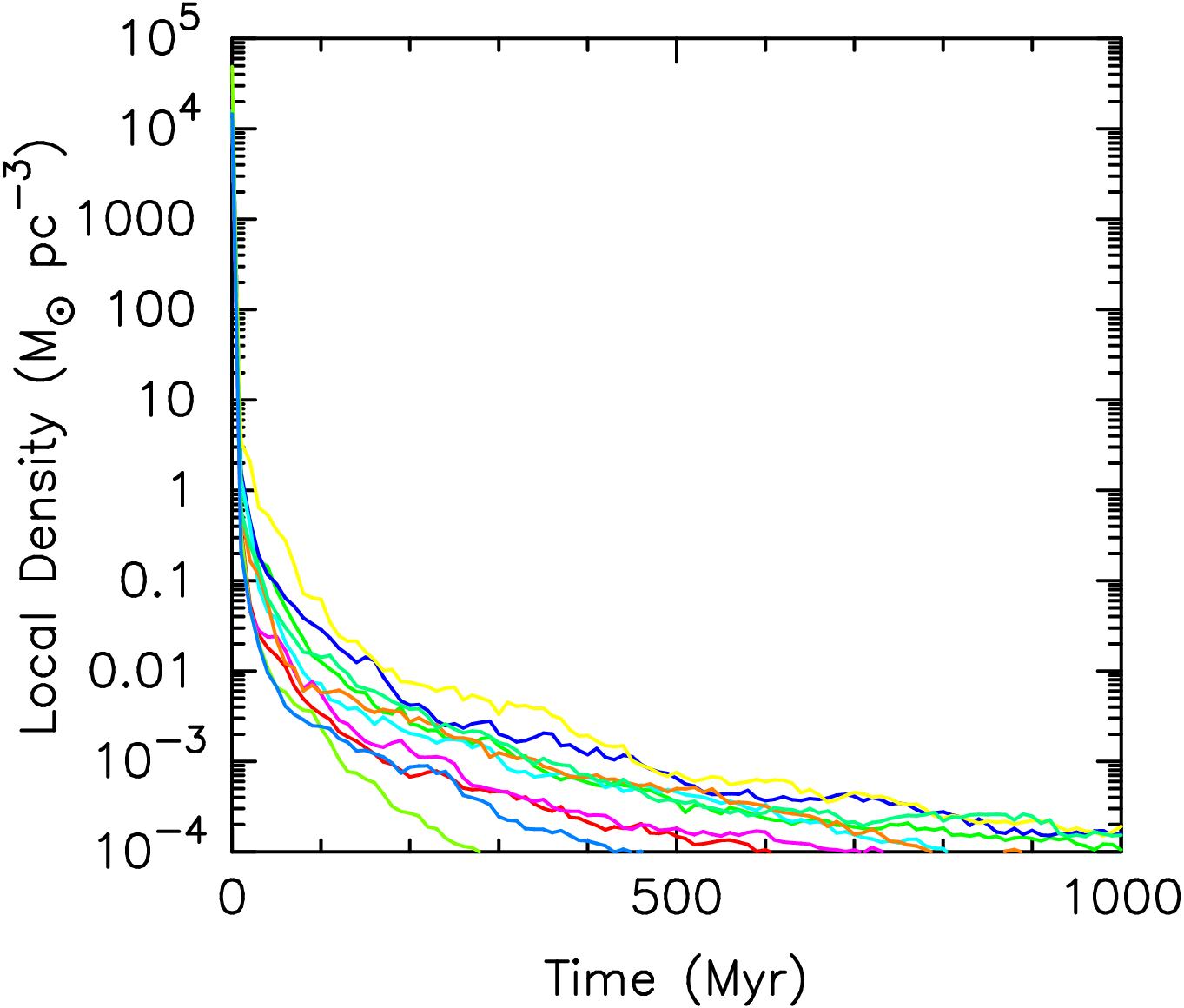}}} 
\hspace*{0.3cm}\subfigure[Plummer sphere]{\label{density_evolution-b}\rotatebox{270}{\includegraphics[scale=0.35]{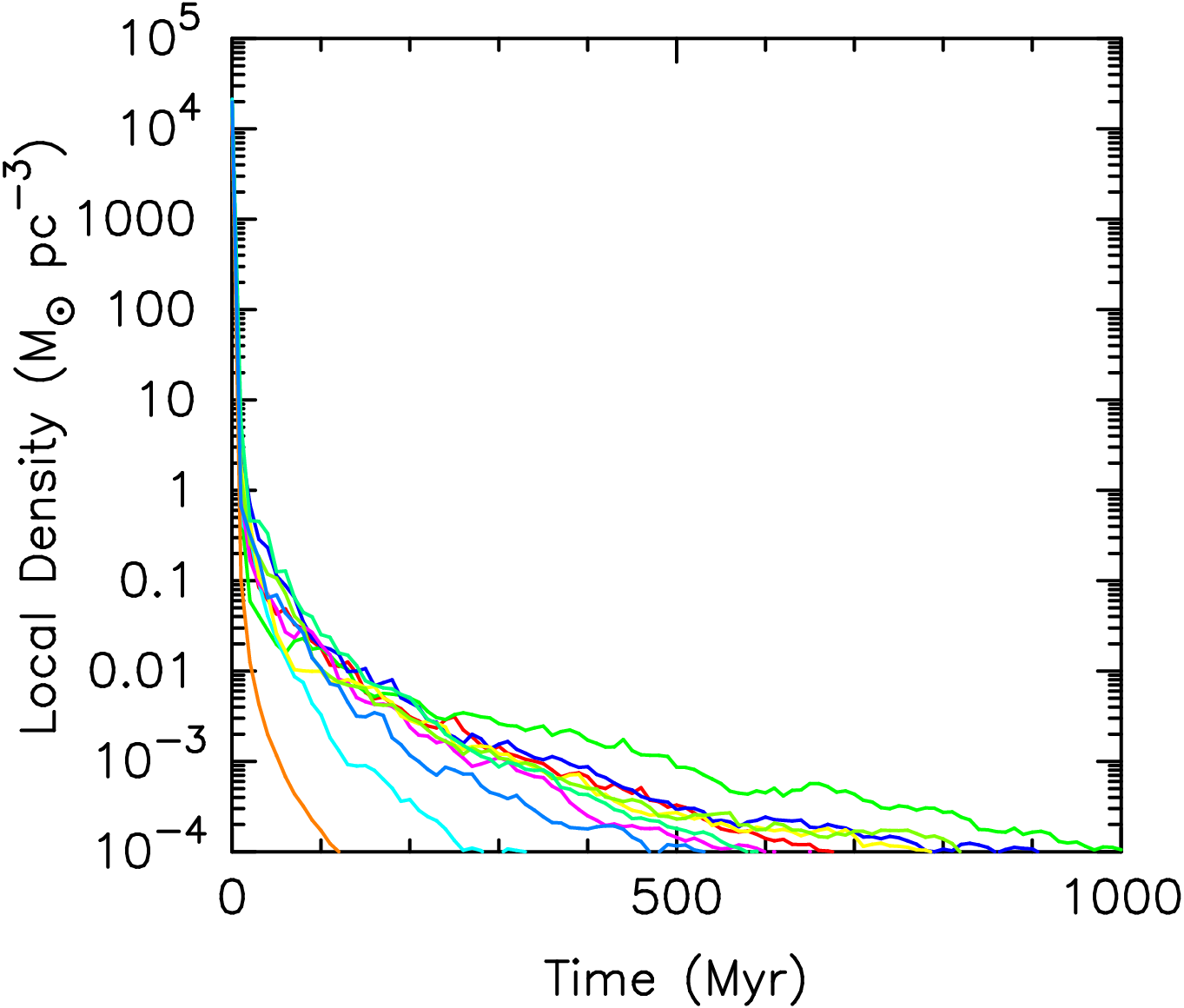}}}

\caption[bf]{Evolution of the median local stellar density in a dense fractal simulation (`A', shown in panel (a)) and in a dense Plummer sphere simulation (`A-Pl', shown in panel (b)). The dynamical evolution of both types of stellar environments is qualitatively similar.}
\label{density_evolution}
  \end{center}
\end{figure*}

\begin{figure}
\begin{center}
\rotatebox{270}{\includegraphics[scale=0.35]{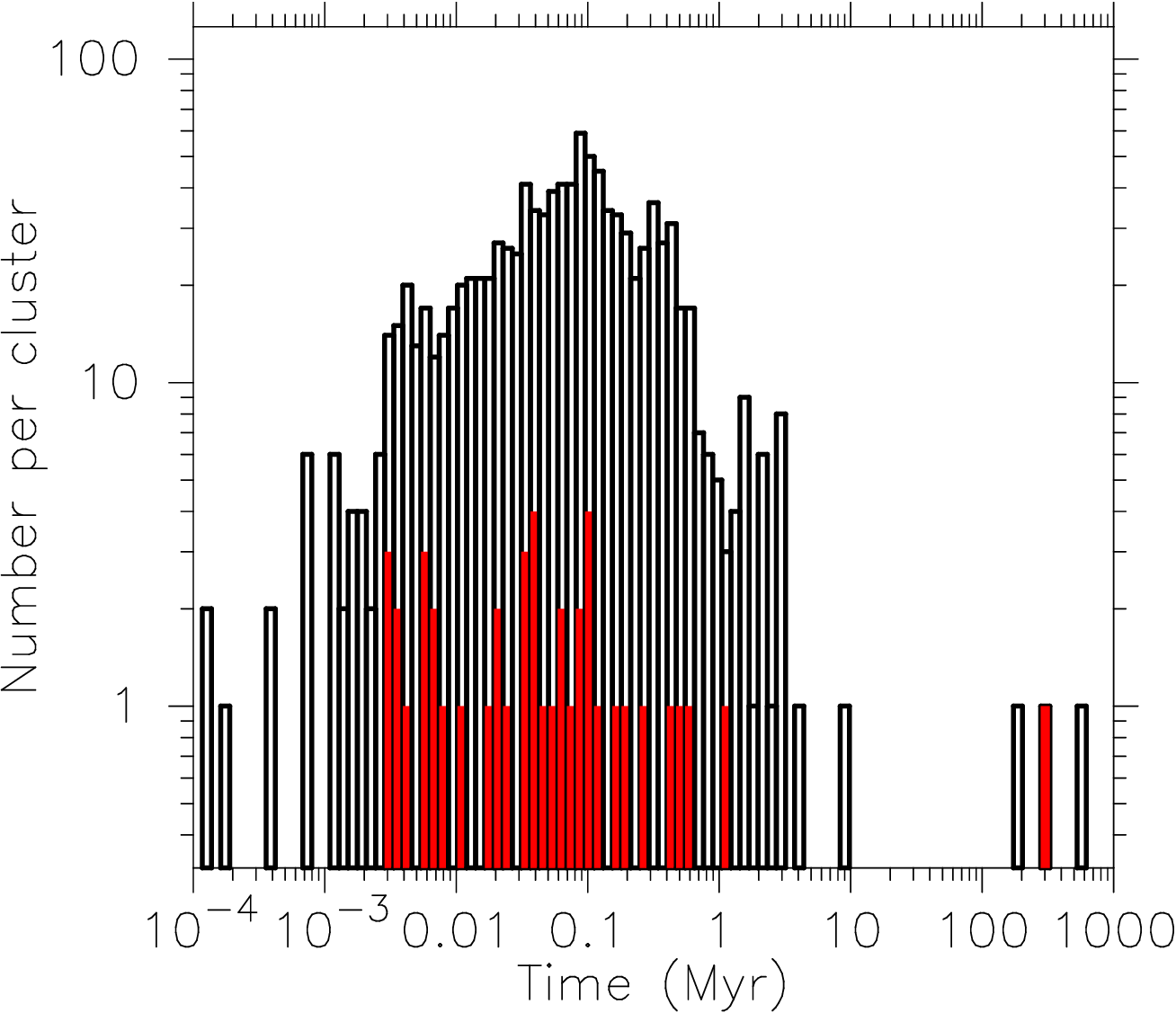}}
\end{center}
\caption[bf]{The number of encounters over time in a representative simulation from one of the Plummer sphere models (set `A-Pl'). The open histogram shows the number of encounters for all stars, and the red histogram shows the numbers of stars that will become white dwarfs after leaving the main sequence. }
\label{Plummer_encounters}
\end{figure}

We follow the dynamical evolution of our star-forming regions for 1\,Gyr, during which time the initial substructure within the fractal distribution is quickly erased (during the first 10\,Myr) and the star-forming region evolves to a dense, centrally concentrated cluster. This later evolution is indistinguishable from the evolution of a Plummer sphere, though the fractal initially undergoes violent relaxation. However, the evolution of the  median stellar density, shown in Fig.~\ref{density_evolution}, is indistinguishable, leading to similar encounter rate histories between fractal and Plummer sphere simulations (contrast the fractal encounter history shown in Fig.~\ref{fractal_encounters} with the corresponding plot for a Plummer sphere, shown in Fig.~\ref{Plummer_encounters}).

The two sets of simulations shown in Fig.~\ref{density_evolution} have similar initial median stellar densities ($\sim10^4$\,M$_\odot$\,pc$^{-3}$), but the fractals are set up with initial spatial and kinematic substructure to mimic the observations of star-forming regions. The kinematic substructure in the fractals means more planetary systems can form via capture, which typically are on wide orbits. Secondly, the Plummer spheres' lack of substructure, and initially denser central cores, means that fewer of the wider systems survive. Therefore, there are modest differences in the evolution of the planets' semimajor axes distributions between the fractal and Plummer initial conditions.

\begin{figure*}
  \begin{center}
\setlength{\subfigcapskip}{10pt}
\hspace*{-1.5cm} \subfigure[0.1 -- 50\,au]{\label{Plummer_sma-a}\rotatebox{270}{\includegraphics[scale=0.35]{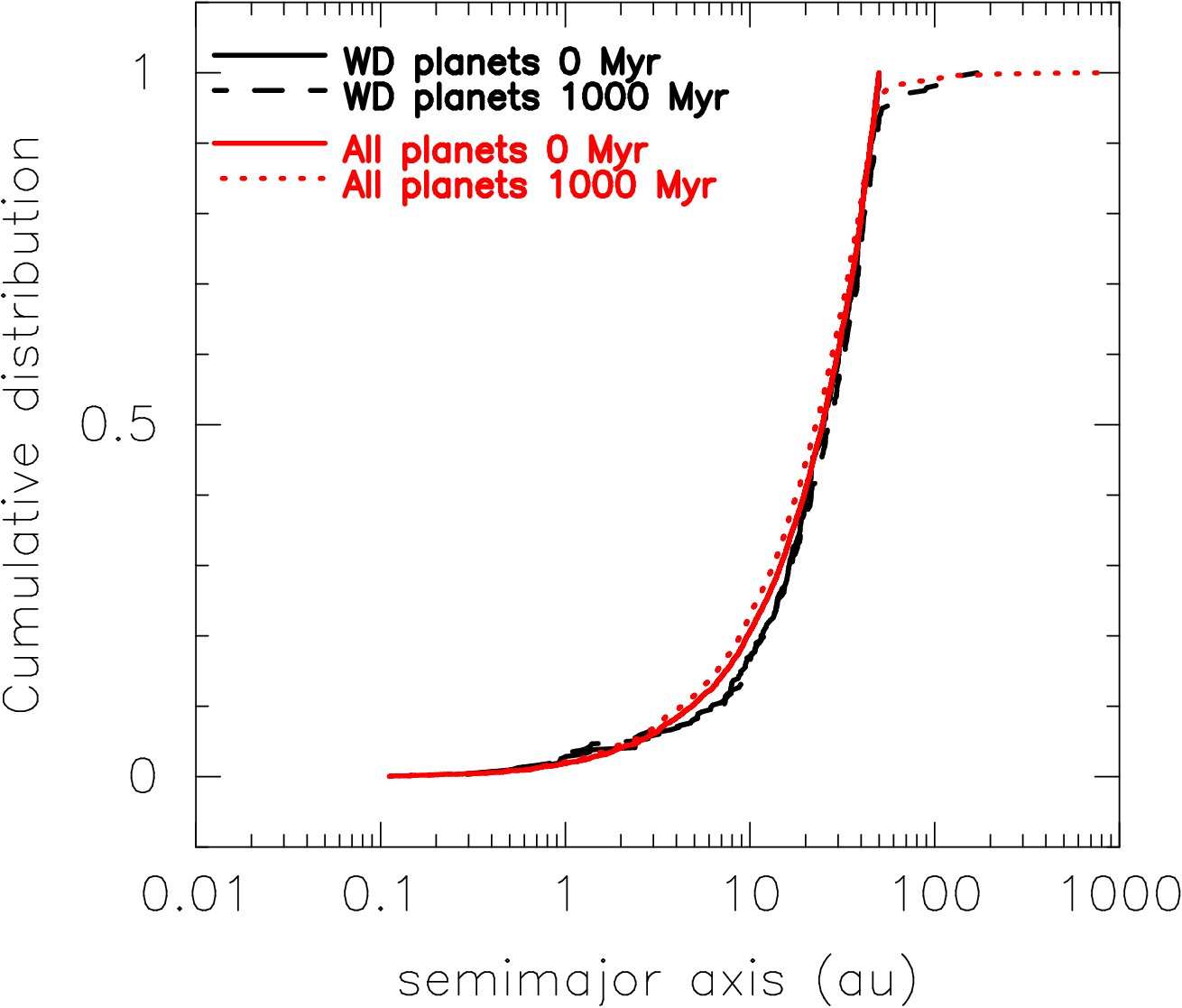}}} 
\hspace*{0.3cm}\subfigure[30\,au]{\label{Plummer_sma-b}\rotatebox{270}{\includegraphics[scale=0.35]{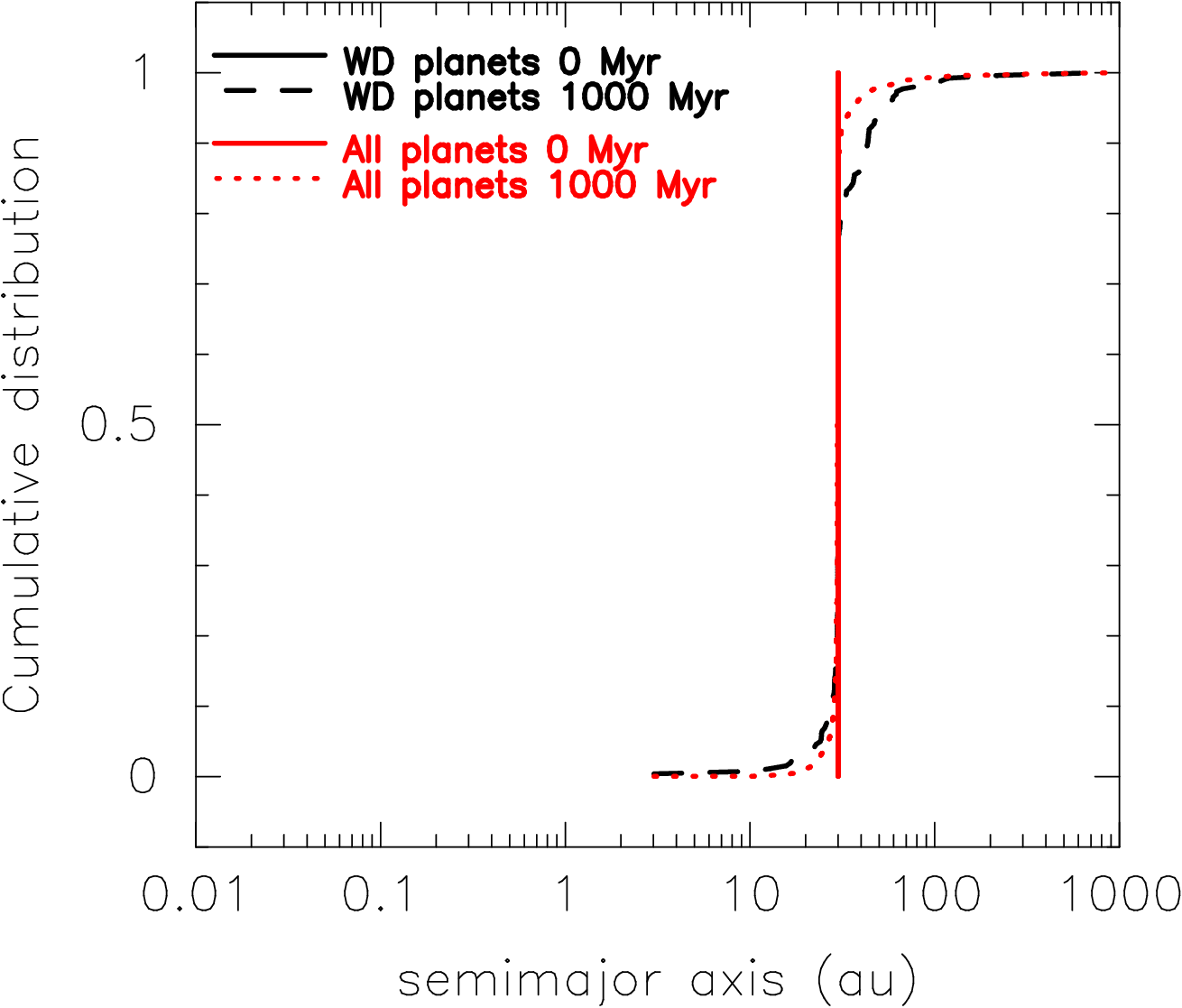}}}

\caption[bf]{Planetary semimajor axis distributions in dense Plummer sphere clusters. Panel (a) shows simulations where the semimajor axes of planets are drawn from a flat distribution between 0.1 -- 50\,au (simulation set `A-Pl'), and panel (b) shows simulations where the semimajor axes are drawn from a delta function at 30\,au (simulation set `E-Pl').}
\label{Plummer_sma}
  \end{center}
\end{figure*}

We show this in Fig.~\ref{Plummer_sma}(a), where the initial semimajor axis distributions for all planets are skewed to smaller separations compared  to those around white dwarf progenitors, likely due to the higher binding energies of the latter systems. When we place all planets at 30\,au (Fig.~\ref{Plummer_sma}(b)) there is no difference between the two initial distributions.

In the simulations with planets placed at 0.1 -- 50\,au (Fig.~\ref{Plummer_sma}(a)), the fraction of systems whose semimajor axes are increased by stellar evolution is around 5\,per cent in the Plummer spheres compared to 10\,per cent in the fractals. In the simulations where the planets are all placed at 30\,au (Fig.~\ref{Plummer_sma}(b)), the difference is similar, with  20\,per cent of planets around white dwarf progenitors affected by stellar evolution in both the fractals and the Plummer spheres.

Apart from forming more systems via capture in the fractals, this suggests that differences between the long-term evolution of fractals and the long-term evolution of Plummer spheres are not important for the the focus of this particular study.

\label{lastpage}

\end{document}